%% file: NonGKSGDenoiserR1.tex
\documentclass[journal]{IEEEtran}
\ifCLASSINFOpdf
\else
\fi
\usepackage{graphicx}
\input{defs.tex}

\usepackage{color}

\usepackage{silence}
\usepackage{soul}
\theoremstyle{plain} 
\usepackage{newtxtext}
\newtheorem{thm}{Theorem}[section] 
\newcommand{\thistheoremname}{}
\newtheorem{genericthm}[thm]{\thistheoremname}

\newtheorem*{genericthm*}{\thistheoremname}
\newenvironment{namedthm*}[1]
  {\renewcommand{\thistheoremname}{#1}%
   \begin{genericthm*}}
  {\end{genericthm*}}
  
\usepackage{diagbox}
\usepackage{booktabs}
\usepackage[T1]{fontenc}
 
\usepackage{mathtools, cuted}
\usepackage{relsize}
\usepackage[export]{adjustbox}
\usepackage[dvipsnames]{xcolor}
\usepackage[skins]{tcolorbox}

\definecolor{darkred}{rgb}{0.6,0,0}
\definecolor{darkgreen}{rgb}{0,0.5,0}
\definecolor{darkblue}{rgb}{0,0,0.5}
\definecolor{goldenrod}{rgb}{0.85, 0.65, 0.13}
\definecolor{goldenbrown}{rgb}{0.6, 0.4, 0.08}
\usepackage{tikz,pgfplots}
\pgfplotsset{compat=1.5.1}
\usetikzlibrary{intersections, pgfplots.fillbetween}
\usepgfplotslibrary{groupplots}
\newcommand{\stepsize}{\alpha}

\usepackage{multicol}

\newcommand{\MRcb}[1]{{\color{black}#1}}

\newcommand{\PropM}{GKSM\xspace}
\newcommand{\KSMs}{KSMs\xspace}
\newcommand{\KSM}{KSM\xspace}
\long\def\red#1{\bgroup\color{red}#1\egroup}

\begin{document}

\title{A Convergent Generalized Krylov Subspace Method 
\\for Compressed Sensing MRI Reconstruction 
\\with Gradient-Driven Denoisers
}


\author{Tao Hong \IEEEmembership{Member, IEEE}, Umberto Villa \IEEEmembership{Member, IEEE}, and Jeffrey A. Fessler \IEEEmembership{Fellow, IEEE}





\thanks{T. Hong and U. Villa are with the Oden Institute for Computational Engineering and Sciences,
University of Texas at Austin, Austin, TX 78712, USA.
(Email: \texttt{\{tao.hong,uvilla\}@austin.utexas.edu}).
TH and UV were supported in part by NIH grant R01 EB034261.
}

\thanks{J. Fessler
 is with the Department of Electrical and Computer Engineering,
 University of Michigan, Ann Arbor, MI 48109, USA 
 (Email: \texttt{fessler@umich.edu}). Supported in part by NIH grants
R01 EB035618 and R21 EB034344.}

}


%
%

\markboth{}
{Shell \MakeLowercase{\mrmit{et al.}}: Bare Demo of IEEEtran.cls for IEEE Journals}
%


\maketitle

\begin{abstract}
Model-based reconstruction plays a key role in compressed sensing (CS) MRI,
as it incorporates effective image regularizers
to improve the quality of reconstruction.
The Plug-and-Play and Regularization-by-Denoising frameworks leverage advanced denoisers
(e.g., convolutional neural network (CNN)-based denoisers)
and have demonstrated strong empirical performance.
However, their theoretical guarantees remain limited,
as practical CNNs often violate key assumptions.
In contrast, gradient-driven denoisers achieve competitive performance,
and the required assumptions for theoretical analysis are easily satisfied.
However, solving the associated optimization problem remains computationally demanding.
To address this challenge,
we propose a generalized Krylov subspace method (\PropM)
to solve the optimization problem efficiently.
Moreover, we also establish rigorous convergence guarantees
for \PropM in nonconvex settings.
Numerical experiments on CS MRI reconstruction 
with spiral and radial acquisitions
validate both the computational efficiency of \PropM
and the accuracy of the theoretical predictions.
The proposed optimization method is applicable
to any linear inverse problem.
\end{abstract}


%


\IEEEpeerreviewmaketitle

\section{Introduction}
\label{sec:Introduction}
%

\IEEEPARstart{M}{AGNETIC} resonance imaging (MRI) scanners acquire k-space data that represents the Fourier coefficients of the image of interest. However, the acquisition process is inherently slow due to physical, hardware, and sampling constraints~\cite{brown2014magnetic}. This slow acquisition presents several practical challenges, including
patient discomfort, motion artifacts, and reduced throughput.
Since the seminal work in~\cite{lustig2007sparse}, compressed sensing (CS)  MRI has attracted significant attention in the MRI community \cite{lustig2008compressed,fessler2010model} for accelerating the acquisition process through structured sampling patterns.
Modern CS MRI methods incorporate multiple receiver coils (a.k.a. parallel imaging~\cite{pruessmann1999sense,griswold2002generalized}) to further improve acquisition speed. Image reconstruction in CS MRI requires solving the following composite minimization problem~\MRcb{\cite{block2007undersampled}}:
\begin{equation}
\label{eq:CSMRIReco:origP}
\uvx^* = \arg\min_{\uvx\in\mathbb{C}^N} F(\uvx) \triangleq \underbrace{\frac{1}{2}\|\umA\uvx - \uvy\|_2^2}_{h(\uvx)} + \lambda\, f(\uvx),
\end{equation}
where $\umA \in \mathbb{C}^{MC \times N}$ denotes the forward operator that maps the image $\uvx \in \mathbb{C}^N$ to the measured k-space data $\uvy \in \mathbb{C}^{MC}$. Here, we consider $C$ receiver coils. The encoding operator $\umA$ is a stack of $C$ submatrices $\umA_c\in\mathbb C^{M\times N}$, each defined as $\umA_c = \umP \umF \umS_c$, where $\umP$ is the sampling mask, $\umF$ is the (non-uniform) Fourier transform, and $\umS_c$ is the coil sensitivity map corresponding to the $c$th coil, which is patient-dependent. The trade-off parameter $\lambda>0$ balances $h(\uvx)$ and $f(\uvx)$. 

The data-fidelity term $h(\uvx)$ in~\eqref{eq:CSMRIReco:origP} promotes consistency with the acquired k-space data. In practice, often $M\ll N$ due to under-sampling, making~\eqref{eq:CSMRIReco:origP} ill-posed. Therefore, incorporating prior knowledge through the regularizer $f(\uvx)$ is essential for stabilizing the reconstruction. The choice of regularization plays a crucial role in reconstruction quality. Traditional hand-crafted regularizers include wavelets~\cite{guerquin2011fast}, total variation (TV)~\MRcb{\cite{rudin1992nonlinear,block2007undersampled}}, combinations of wavelets and TV~\cite{lustig2007sparse,hong2024complex}, dictionary learning~\cite{aharon2006k,ravishankar2011mr}, and low-rank models~\cite{dong2014compressive}, to name a few.
For reviews of various choices for $f(\uvx)$, see~\cite{fessler2010model, fessler2020optimization,ravishankar2019image}.

In the past decade, deep learning (DL) has gained prominence in MRI reconstruction due to its capacity to learn complex image priors directly from large training  datasets~\cite{heckel2024deep}. Roughly speaking, DL-based approaches can be broadly categorized into end-to-end networks~\cite{Wang2016.etal} and physics-driven unrolled algorithms~\cite{aggarwal2018modl,gilton2021deep,ramzi2022nc}. Recently, generative models have emerged as a powerful class for modeling priors in MRI, achieving impressive results across various settings~\cite{chung2022score}. 

An alternative to classical DL pipelines is the Plug-and-Play (PnP)~\cite{venkatakrishnan2013plug} and REgularization-by-Denoising (RED)~\cite{romano2017little} frameworks. \MRcb{PnP and RED integrate powerful denoisers into iterative reconstruction algorithms and have demonstrated competitive performance across various imaging tasks~\cite{sreehari2016plug,Meinhardt.etal2017,Buzzard.etal2018,hong2019acceleration,zhang2021plug,hong2024provable}. Early PnP/RED frameworks employed classical denoisers such as the median filter~\cite{huang1979fast}, non-local means~\cite{buades2005non}, and BM3D~\cite{dabov2007image}. As convolutional neural network (CNN)–based denoisers have shown superior performance to classical ones, modern PnP/RED methods typically integrate CNN-based denoisers.} Unlike end-to-end or unrolled DL methods that require retraining for each imaging task, PnP and RED leverage learned image priors to flexibly adapt to changes in the forward model without retraining. This adaptability is particularly beneficial in CS MRI reconstruction, where scan-specific variations (e.g., different sampling trajectories and patient-specific sensitivity maps) are common.
See~\cite{ahmad2020plug} for a review of PnP methods \MRcb{that incorporate both classical and CNN-based denoisers} in MRI reconstruction. 

Despite the empirical success of PnP and RED, their theoretical convergence guarantees remain an active area of research; see~\cite{chan2017plug,reehorst2018regularization,Buzzard.etal2018,ryu2019plug,terris2020building,kamilov2023plug}.
These works typically require that the denoisers either approximate maximum a posteriori or minimum mean squared error estimators, or satisfy the nonexpansiveness condition. However, many successful denoisers—especially those based on CNNs—do not satisfy these assumptions. As a result, PnP and RED with such denoisers cannot be rigorously interpreted as optimization algorithms. Although optimization-free perspectives have been proposed, understanding the behavior of these frameworks remains challenging~\cite{Buzzard.etal2018}. One alternative is to train denoisers with additional regularization that enforces a bounded Lipschitz constant~\cite{ryu2019plug,terris2020building}. However, guaranteeing strict and tight boundedness in practice remains an open challenge.

Recent efforts have aimed to close the gap between the theoretical foundations and practical effectiveness of PnP and RED by introducing gradient-driven denoisers~\cite{cohen2021has, hurault2021gradient, chaudhari2024gradient}. In this approach, the unknown image $\uvx$ is recovered by solving 
\begin{equation}
\label{eq:CSMRIReco:rewrite}
    \uvx^* = \arg\min_{\uvx \in \mathcal C} \, F(\uvx) \coloneqq \underbrace{\frac{1}{2} \|\umA \uvx - \uvy\|_2^2}_{h(\uvx)} + \lambda f_{\bm\uptheta}(\uvx),
\end{equation}
where $f_{\bm\uptheta}(\uvx)$ is a scalar-valued energy function parameterized by CNNs that serves as a learned image prior and $\mathcal C$ is a closed convex set in $\mathbb{C}^N$.
The parameters $\bm\uptheta$ are learned by enforcing $\uvx - \nabla_{\uvx} f_{\bm\uptheta}(\uvx)$ to act as a denoiser. Thus, the only required assumption is the differentiability of $f_{\bm\uptheta}$ with respect to $\uvx$, which allows one to integrate deep learning into inverse problems while maintaining a degree of interpretability—an essential requirement in medical imaging, where reconstructions directly influence diagnostic decisions. For notational simplicity, we omit the subscript $\bm\uptheta$ and use $\nabla f(\uvx)$ instead of $\nabla_{\uvx} f(\uvx)$. Moreover, we absorb $\lambda$ into $f(\uvx)$ in the following discussion since $\lambda$ is fixed throughout the minimization once it is selected.
 
Although both $h(\uvx)$ and $f(\uvx)$ in~\eqref{eq:CSMRIReco:rewrite} are differentiable, $f(\uvx)$ is generally nonconvex, which poses challenges for designing convergent and efficient algorithms. Cohen et al.~\cite{cohen2021has} applied a projected gradient descent method with a line search to solve~\eqref{eq:CSMRIReco:rewrite}. Alternatively, Hurault et al.~\cite{hurault2021gradient} employed the proximal gradient descent method with line search. Both approaches provide convergence guarantees under the assumption that $\nabla f$ is Lipschitz continuous. However, these methods typically require hundreds of iterations to converge, which limits their practical applicability. Recently, Hong et al.~\cite{hong2025CQNPMCSMRI} proposed a convergent complex quasi-Newton proximal method (CQNPM)
that significantly reduces the computational time required to solve~\eqref{eq:CSMRIReco:rewrite}. Their convergence is established under the assumptions that $\nabla f$ is Lipschitz continuous and that the proximal Polyak-{{\L}}ojasiewicz condition holds. Although CQNPM converges faster than existing methods for solving~\eqref{eq:CSMRIReco:rewrite}, it requires solving a weighted proximal mapping (as defined in~\cite[equation (3)]{hong2025CQNPMCSMRI}) at each iteration.
This step requires computing $\umA\uvx$ and $\umA^\HTrans\uvx$
 multiple times,%
\footnote{$\HTrans$ denotes the Hermitian transpose operator.}
which can increase the overall computational complexity.
Computing $\umA\uvx$ is expensive
in MRI reconstruction with many coils, high-resolution images,
or many interleaves or spokes in non-Cartesian acquisitions.
Drawing inspiration from Krylov subspace methods (KSMs)~\cite{saad2003iterative},
we propose a generalized Krylov subspace method (\PropM)
for efficiently solving~\eqref{eq:CSMRIReco:rewrite},
which requires computing
$\umA\uvx$, $\umA^\HTrans\uvx$, and $\nabla f(\uvx)$ only once per iteration.
Our main contributions are summarized as follows:
\begin{itemize}
    \item We propose a generalized Krylov subspace method (\PropM) for efficiently solving~\eqref{eq:CSMRIReco:rewrite}.
    \item We present a rigorous convergence analysis of \PropM in nonconvex settings, along with the convergence rate of the cost function values.
    \item \MRcb{We extensively evaluate the performance of \PropM on brain (respectively, knee) images from the dataset described in the MoDL paper~\cite{aggarwal2018modl} (respectively, the NYU fastMRI dataset~\cite{zbontar2018fastmri}). The k-space data are simulated from the reconstructed complex-valued images using spiral and radial sampling trajectories. We also empirically validate the accuracy of the convergence analysis.}
\end{itemize}

The rest of this paper is organized as follows.
\Cref{sec:KSInverse} reviews the preliminaries on \KSMs
and discusses related work that generalizes \KSMs for solving inverse problems.
\Cref{sec:ProposedMethod} describes \PropM in detail.
\Cref{sec:ConvAnalysis} provides a rigorous convergence analysis of \PropM.
\Cref{sec:NumericalExps} reports experimental results
that evaluate the performance of \PropM
and empirically validate the theoretical analysis. 


\section{Preliminaries on Krylov Subspace Methods}
\label{sec:KSInverse}
This section first introduces \KSMs, which were primarily developed for solving linear equations. We then review related generalized Krylov methods for linear inverse problems, along with existing theoretical results. Our main goal in this section is to provide a sketch of the key developments in \KSMs, from their origins in solving linear equations to their generalization for inverse problems.

\KSMs are a class of iterative algorithms for solving problems of the form
\begin{equation}
\label{eq:LinearEq}
\bar{\umA} \uvx = \uvb,
\end{equation}
where $\bar{\umA} \in \mathbb{R}^{N \times N}$ is typically sparse, ill-conditioned, and large-scale. At $k$th iteration, \KSMs construct an approximate solution to $\uvx^*$ within the Krylov subspace:
\begin{equation}
\label{eq:KrylovSubspace}
\mathcal{K}_k(\bar{\umA}, \bar{\uvr}_1) = \mathrm{span}\left\{\bar{\uvr}_1, \bar{\umA} \bar{\uvr}_1, \bar{\umA}^2\bar{\uvr}_1, \dots, \bar{\umA}^{k-1} \bar{\uvr}_1\right\},
\end{equation}
where $\bar{\uvr}_1 = \uvb - \bar{\umA} \uvx_1$ is the initial residual. The approximate solution $\mathbf{x}_{k+1}$ is obtained by seeking $\mathbf{x}_{k+1} \in \mathbf{x}_1 + \mathcal{K}_k(\bar{\umA}, \bar{\uvr}_1)$ that minimizes a chosen norm of the residual. The most widely used \KSMs include, but are not limited to, the conjugate gradient method~\cite{hestenes1952methods}, LSQR~\cite{paige1982lsqr}, BiCGSTAB~\cite{van1992bi}, and the generalized minimal residual method~\cite{saad1986gmres}. These methods are designed for different types of matrices $\bar{\umA}$, such as symmetric positive definite, non-symmetric, or indefinite, to name a few~\cite{saad2003iterative}. Moreover, \KSMs can incorporate preconditioners to further accelerate convergence~\cite{saad1993flexible}.

Many inverse problems with variational regularizers~\cite{chung2024computational,hong2025using} can be modeled as the following $\ell_p$-$\ell_q$ optimization problem:
\begin{equation}
\label{eq:lplq}
\min_{\uvx\in\mathbb R^N} \frac{1}{p} \|\umA \uvx - \uvy\|_p^p + \frac{\lambda}{q} \|\umW \uvx\|_q^q,
\end{equation}
where $0 < p, q \leq 2$, and $\umW$ represents a transform such as a wavelet transform.  \KSMs have been generalized to solve problems such as~\eqref{eq:lplq}.   \MRcb{For $p=q=2$, Lampe et al. \cite{lampe2012large} presented a generalized \KSM to address~\eqref{eq:lplq} in which the parameter $\lambda$ is adaptively adjusted to keep $\|\umA\uvx-\uvy\|_2$ sufficiently close to a prescribed tolerance.} Lanza et al.~\cite{lanza2015generalized} proposed to solve~\eqref{eq:lplq}
using a \KSM along with an iteratively reweighted approach.
Moreover, Huang et al. combined Krylov subspace-based method with majorization minimization for solving ~\eqref{eq:lplq}~\cite{huang2017majorization}.
To avoid the inner-outer iterations when using the iteratively approach in~\cite{lanza2015generalized},
several flexible Krylov subspaces were proposed to improve  efficiency~\cite{gazzola2014generalized,chung2019flexible,gazzola2021iteratively}. See~\cite{gazzola2020krylov,chung2024computational} for a review of using \KSMs for inverse problems.

Besides their extension to inverse problems, rigorous convergence analyses of \KSMs remain an open research area. Lanza et al.~\cite{lanza2015generalized} showed that the iterates converged to a minimizer
of~\eqref{eq:lplq}
for $1 \leq p, q \leq 2$ if $\mathrm{ker}(\umA^\Trans \umA) \cap \mathrm{ker}(\umW^\Trans \umW) = \{\bm 0\}$, where $\mathrm{ker}(\cdot)$ denotes the null space of the matrix, and the constructed Krylov subspace fully represents the entire image domain. Similar convergence results can also be found in~\cite{huang2017majorization,buccini2020modulus}. Other works~\cite{gazzola2014generalized,chung2019flexible,gazzola2021iteratively} proved that the cost function values are monotonically decreasing and that the iterates converge to a stationary point. For brevity, we present only the convergence results of \KSMs for inverse problems.  See~\cite{sterck2012nonlinear,sterck2021asymptotic} and the references therein for discussions on \KSMs and their convergence in other contexts.

\section{Proposed Method}
\label{sec:ProposedMethod}

This section provides the details of our \PropM for solving~\eqref{eq:CSMRIReco:rewrite}. We first discuss the case where $\mathcal C = \mathbb C^N$ and then describe how to incorporate a convex constraint. Lastly, we provide further discussion of \PropM to offer additional insights.

Given a subspace basis $\umV_k\in\mathbb C^{N\times k}$ satisfying $\umV_k^\HTrans \umV_k=\umI_k$ where $\umI_k$ is the identity matrix with dimension $k$, a Hermitian positive definite matrix $\umB_k\in\mathbb C^{N\times N},~\umB_k\succ 0$, and $\stepsize_k\in\mathbb R,~\stepsize_k>0$, \PropM solves the following problem at the $k$th iteration to obtain the coefficient $\vbeta_k \in \mathbb{C}^k$:
\MRcb{
\begin{equation}
	\label{eq:CSMRIReco:origP:GradDenoiser:MM}
	\vbeta_{k}=\arg\min_{\vbeta\in\mathbb C^k} \underbrace{\frac{1}{2}\|\umA\uvx-\uvy\|_2^2+\bar f(\uvx,\uvx_k,\umB_k,\stepsize_k)}_{\bar F(\uvx,\uvx_k)},
\end{equation}
where $\uvx=\umV_k\vbeta$}, $\bar f(\uvx,\uvx_k,\umB_k,\stepsize_k)\equiv \left\langle \nabla f(\uvx_k),\uvx \right\rangle+\frac{1}{2\stepsize_k}\|\uvx-\uvx_k\|_{\umB_k}^2$
is a quadratic proximal term, and $\|\uvx\|^2_{\umB_k}=\uvx^\HTrans \umB_k \uvx$.
Then the next image iterate is $\uvx_{k+1}=\umV_k\vbeta_k$.
Rewriting~\eqref{eq:CSMRIReco:origP:GradDenoiser:MM} in terms of $\vbeta$ and reorganizing yields 
\begin{equation}
\label{eq:CSMRIReco:origP:GradDenoiser:MM:Beta}
	\vbeta_k=\arg\min_{\vbeta\in\mathbb C^k} \left \|\bmat \umA\umV_k\vbeta \\ \bar{\umB}_k^{\frac{1}{2}}\umV_k\vbeta \emat -  \bmat \uvy \\  \bar{\umB}_k^{\frac{1}{2}}\uvw_k \emat \right \|_2^2,
\end{equation}
where $\uvw_k=\uvx_k-\stepsize_k\umB_k^{-1}\nabla f(\uvx_k)$ and \MRcb{$\bar{\umB}_k^{1/2}$ denotes the principal matrix square root of 
$\bar{\umB}_k = \umB_k/\stepsize_k$ and it is unique \cite{fessler2024linear}}. \MRcb{Note that $\umA\umV_k$ is built incrementally during the algorithm.} Compared with the image size, the dimension of $\vbeta_k$ is relatively low \MRcb{because the number of iterations will be significantly smaller than the image dimension.
Therefore,} we solve~\eqref{eq:CSMRIReco:origP:GradDenoiser:MM:Beta} directly, i.e., 
\begin{equation}
\label{eq:CSMRIReco:origP:GradDenoiser:ComputeBeta}
	\vbeta_k=(\umV_k^\HTrans\umA^\HTrans\umA\umV_k+\umV_k^\HTrans\bar{\umB} _k\umV_k)^{-1}\umV_k^\HTrans(\umA^\HTrans\uvy+\bar{\umB}_k\uvw_k).
\end{equation}
Here the matrix being inverted is only $k \times k$
with $k \ll N$.

To enrich the subspace after the $k$th iteration,
we first compute the gradient of the objective function in~\eqref{eq:CSMRIReco:origP:GradDenoiser:MM} with respect to $\uvx$ at $\uvx=\uvx_{k+1}$, i.e.,
\begin{equation}
\label{eq:ComputeResidual}
\uvr_k = \nabla_{\uvx} \bar F(\uvx_{k+1},\uvx_k). 
\end{equation}
Then we set $\uvv_{k+1}=\tilde{\uvr}_k/\|\tilde {\uvr}_k\|$ with $\tilde{\uvr}_k=(\umI_N-\umV_k\umV_k^\HTrans)\uvr_k$.
The new subspace basis $\umV_{k+1}$ is formulated as
$\bmat \umV_k  & \uvv_{k+1}\emat$.
If $\|\tilde{\uvr}_k\|= 0$, we simply skip the update of $\umV_k$. \MRcb{Note that the dimension of $\vbeta_k$ is equal to the number of columns 
of $\mathbf{V}_k$. This dimension may be smaller than $k$ if the event 
$\|\tilde{\mathbf{r}}_k\| = 0$ occurs, in which case the number of columns of 
$\mathbf{V}_k$ is also smaller than $k$.}
 \Cref{alg:NGKS} summarizes the detailed steps of \PropM.
To establish the convergence rate of the cost function values, we introduce an additional step~\ref{alg:NGKS:V_k:I} in \Cref{alg:NGKS}, as the generated  $\umV_k$
does not necessarily span the entire image domain.
Note that \PropM reduces to CQNPM~\cite{hong2025CQNPMCSMRI} when $\umV_k=\umI_N$.
For this case, we simply apply the accelerated gradient descent method
to solve~\eqref{eq:CSMRIReco:origP:GradDenoiser:MM}.
If the regularizer $f$ in~\eqref{eq:CSMRIReco:rewrite}
is a quadratic function,
then the subspace spanned by $\umV_k$ simplifies to the classical Krylov subspace in~\eqref{eq:KrylovSubspace}.
To obtain $\umB_k$, we adopt the algorithm presented
in \cite[Algorithm 2]{hong2025CQNPMCSMRI}
such that $\umB_k$ is an estimate of the Hessian matrix of $f(\uvx)$
that is guaranteed to be Hermitian positive definite.
For completeness,
\Cref{alg:ModfiedZero:SR1} provides the detailed steps for computing $\umB_k$.
The operator $\Re(\cdot)$ in~\Cref{alg:ModfiedZero:SR1} extracts the real part.

\begin{algorithm}[t]        
\caption{Generalized Krylov Subspace Method (\PropM)}    
\label{alg:NGKS} 
\begin{algorithmic}[1]
\REQUIRE $\uvx_1$, stepsize $\stepsize_k > 0$, $\umV_1 = \frac{\umA^\HTrans\uvy}{\|\umA^\HTrans\uvy\|}$, $\umA\umV_1$, maximal number of subspace iterations $K$, and maximal number of total iterations \texttt{Max\_Iter} 
%
\ENSURE 
\FOR {$k=1,2,\dots,\texttt{Max\_Iter}$}
\STATE Compute $\nabla f(\uvx_k)$
\STATE Set $\umB_k$ using~\Cref{alg:ModfiedZero:SR1}
\STATE Compute $\vbeta_k$ using~\eqref{eq:CSMRIReco:origP:GradDenoiser:ComputeBeta} (or solve~\eqref{eq:CSMRIReco:consP:GradDenoiser:MM:Beta} for $\vbeta_k$ if a convex constraint is enforced) \label{alg:computebeta}
\STATE Compute $\uvx_{k+1}\leftarrow \umV_k\vbeta_k$ 
\IF{$k\leq K$}
\STATE Compute $\uvr_k \leftarrow \nabla_{\uvx} \bar F(\uvx_{k+1},\uvx_k)$
\STATE $\tilde{\uvr}_k\leftarrow (\umI-\umV_k\umV_k^\HTrans)\uvr_k$
\IF{$\|\tilde{\uvr}_k\|\neq 0$}
\STATE $\uvv_{k+1}\leftarrow \tilde{\uvr}_k/\|\tilde {\uvr}_k\|$
\STATE $\umV_{k+1}\leftarrow \bmat \umV_k & \uvv_{k+1} \emat$ 
\STATE $\umA\umV_{k+1}\leftarrow \bmat \umA\umV_k & \umA\uvv_{k+1} \emat$ 
\ELSE 
\STATE $\umV_{k+1}\leftarrow \umV_k$
\STATE $\umA\umV_{k+1}\leftarrow \umA\umV_k$ 
\ENDIF
\ELSE
\STATE $\umV_{k+1} \leftarrow \umI_N$\label{alg:NGKS:V_k:I}
\ENDIF
\ENDFOR
\end{algorithmic}
\end{algorithm}

\begin{algorithm}[t]        
\caption{Modified Memory Efficient Self-Scaling Hermitian Rank-$1$ Method}
\label{alg:ModfiedZero:SR1} 
\begin{algorithmic}[1]
\REQUIRE $\uvx_{k-1}$, $\uvx_k$, $\nabla f(\uvx_{k-1})$, $\nabla f(\uvx_k)$, $\delta>0$, $\nu_1\in (0,1)$, and $\nu_2\in(1,\infty)$
\STATE Set $\uvs_k\leftarrow \uvx_k-\uvx_{k-1}$ and $\uvm_k\leftarrow (\nabla f(\uvx_k)-\nabla f(\uvx_{k-1}))$
\STATE Compute $a$ such that 
\begin{equation}
\label{eq:alg:Zero:SC:SR1:alpha}
\begin{array}{c}
     \min_a\{a\in[0,1]|{\bar\uvm}_k=a\uvs_k+(1-a)\uvm_k\}  \\
     \text{satisfies}~\nu_1\leq \frac{\Re(\langle\uvs_k,{\bar\uvm}_k\rangle)}{\langle\uvs_k,\uvs_k\rangle}~\text{and}~\frac{\langle{\bar\uvm}_k,{\bar\uvm}_k\rangle}{\Re(\langle\uvs_k,{\bar\uvm}_k\rangle)}\leq\nu_2 
\end{array}    
\end{equation} \label{alg:ModfiedZero:SR1:stepv_k}
\STATE Compute $\tau_k \leftarrow \frac{\langle\uvs_k,\uvs_k\rangle}{\Re(\langle\uvs_k,{\bar\uvm}_k\rangle)}-\sqrt{\left(\frac{\langle\uvs_k,\uvs_k\rangle}{\Re(\langle\uvs_k,{\bar\uvm}_k\rangle)}\right)^2-\frac{\langle\uvs_k,\uvs_k\rangle}{\langle{\bar\uvm}_k,{\bar\uvm}_k\rangle}}$\label{alg:ModfiedZero:SR1:steptau_k}\\[5pt]
\STATE $\rho_k\leftarrow \Re(\langle \uvs_k-\tau_k{\bar\uvm}_k,{\bar\uvm}_k\rangle)$ \label{alg:ModfiedZero:SR1:stepInnerrhok}
\IF{$\rho_k \leq \delta \|\uvs_k-\tau_k{\bar\uvm}_k\|\|{\bar\uvm}_k\|$ \label{alg:ModfiedZero:SR1:u_k0}}
\STATE $\uvu_k\leftarrow \bm 0$
\ELSE
\STATE $\uvu_k\leftarrow \uvs_k-\tau_k{\bar\uvm}_k$
\ENDIF
\STATE $\rho_k^{\umB} \leftarrow \tau_k^2 \rho_k + \tau_k \uvu_k^\HTrans \uvu_k$
\STATE {\bf Return:~}$
\umB_k \leftarrow \tau_k^{-1} \umI_N - \frac{\uvu_k \uvu_k^\HTrans}{\rho_k^{\umB}}
$
\end{algorithmic}
\end{algorithm}

\subsection{Incorporating A Convex Constraint}
This part extends \PropM to handle a convex constraint
on $\uvx$
with a slight increase in computational cost.
To ensure $\uvx\in\mathcal C$,
we solve the following problem for $\vbeta_k$ instead of~\MRcb{\eqref{eq:CSMRIReco:origP:GradDenoiser:MM:Beta}}:
\begin{equation}
\label{eq:CSMRIReco:consP:GradDenoiser:MM:Beta}
	\vbeta_k=\arg\min_{(\umV_k \vbeta) \in\mathcal C} \left \|\bmat \umA\umV_k\vbeta \\ \bar{\umB}_k^{\frac{1}{2}}\umV_k\vbeta \emat -  \bmat \uvy \\  \bar{\umB}_k^{\frac{1}{2}}\uvw_k \emat \right \|_2^2.
\end{equation}
Letting $\uvz=\umV_k\vbeta$ and using the fact that $\umV_k^\HTrans\umV_k=\umI_k$, we have $\vbeta = \umV_k^\HTrans \uvz$. Thus we rewrite~\eqref{eq:CSMRIReco:consP:GradDenoiser:MM:Beta} as
\begin{equation}
	\label{eq:CSMRIReco:consP:GradDenoiser:MM:Z}
	\uvx_{k+1}=\arg\min_{\uvz\in\mathcal C} \left \|\bmat \umA\umV_k\umV_k^\HTrans \uvz \\ \bar{\umB}_k^{\frac{1}{2}}\umV_k \umV_k^\HTrans\uvz \emat -  \bmat \uvy \\  \bar{\umB}_k^{\frac{1}{2}}\uvw_k \emat \right \|_2^2.
\end{equation}
Since the objective function of~\eqref{eq:CSMRIReco:consP:GradDenoiser:MM:Z} is differentiable, we simply apply the accelerated projection gradient method~\cite{beck2009fast} to solve~\eqref{eq:CSMRIReco:consP:GradDenoiser:MM:Z} efficiently.
Although $\uvz$ has the same dimension as $\uvx$, solving~\eqref{eq:CSMRIReco:consP:GradDenoiser:MM:Z} only requires simple matrix-vector multiplications since $\umA\umV_k$ and $\bar{\umB}_k^{\frac{1}{2}}\umV_k$ are precomputed and saved. In practice, to extend \Cref{alg:NGKS} to handle the convex constraint, we only need to replace the computation of $\vbeta_k$ at step~\ref{alg:computebeta} with the solution of~\eqref{eq:CSMRIReco:consP:GradDenoiser:MM:Beta} by solving~\eqref{eq:CSMRIReco:consP:GradDenoiser:MM:Z}.





\subsection{Discussion}
\label{sec:ProposedMethod:sub:Dis}
The dominant computations at each iteration in~\Cref{alg:NGKS} involve computing $\nabla f(\uvx)$, $\umA\uvx$ and $\umA^\HTrans\uvx$ once,
and the overall computational cost per iteration is lower than that of the methods proposed in~\cite{hurault2021gradient,hong2025CQNPMCSMRI}
that require dozens of evaluations of $\umA\uvx$ and $\umA^\HTrans\uvx$ per iteration.
Apart from computational efficiency, \PropM requires additional memory,
because it must store $\umV_k$ and $\umA\umV_k$.
Thus, \PropM may become memory-prohibitive for very large-scale problems.
A practical heuristic to address this challenge is to use a restart strategy, in which we cyclically set $\umV_k = \uvx_{k+1}/\|\uvx_{k+1}\|$. This not only reduces the memory usage but also lowers the computational cost. \MRcb{Algorithmically, when a restart is triggered, this is equivalent to rerunning 
\Cref{alg:NGKS} with $\uvx_{k+1}$ as the new initial value and  $\umV_1 = \uvx_{k+1} / \|\uvx_{k+1}\|$.} Moreover, in practice, there is no guarantee that a column-orthogonal matrix $\umV_{k+1}$ can always be constructed from $\uvr_k$, since $\|\tilde{\uvr}_k\|$ may be zero. The restart strategy typically helps to escape such a situation. However, in our experimental settings, we never found that $\|\tilde{\uvr}_k\|=  0$.
We leave the study of restart strategies to future work.

The following convergence analysis shows that \PropM is guaranteed to monotonically decrease the cost function value every iteration.
Thus, it is safe to set $K = \texttt{Max\_Iter}$ in practice. However, the convergence rate of the cost function values to a minimum remains unclear since we cannot guarantee that $\umV_k$ will span the entire image space after a finite number of iterations. To better characterize the convergence rate of the cost values, we introduce step~\ref{alg:NGKS:V_k:I} in~\Cref{alg:NGKS}. This addition allows us to explicitly quantify the cost convergence after $K$ iterations.

\section{Convergence Analysis}
\label{sec:ConvAnalysis}

This section provides a rigorous convergence analysis of using~\Cref{alg:NGKS} to solve~\eqref{eq:CSMRIReco:rewrite}.
Because the unconstrained problem is a special case of the constrained one, 
we focus our analysis on using \PropM for problems with constraints.
We use the notation $F_{\mathcal{C}}(\uvx) = F(\uvx) + \iota_{\mathcal{C}}(\uvx)$ in the following analysis,
where $ \iota_{\mathcal{C}}(\uvx)$ denotes the characteristic function, defined as $\iota_{\mathcal{C}}(\uvx) = 0$ if $\uvx \in \mathcal{C}$, and $\iota_{\mathcal{C}}(\uvx) = +\infty$ otherwise. Here, we assume that $\mathcal{C}$ is convex and that its indicator function $\iota_{\mathcal{C}}$ is lower semicontinuous. Before presenting our main convergence results, we first review the definition of Kurdyka–{\L}ojasiewicz (KL) inequality and make one assumption, followed by four supporting lemmas.

\begin{definition}[Kurdyka–{\L}ojasiewicz inequality \cite{attouch2010proximal,bolte2014proximal}]
\label{def:KLIneq}
Let $\chi(\uvx): \mathbb{C}^N \to (-\infty, +\infty]$ be a proper, lower semicontinuous function. We say that $\chi$ satisfies the Kurdyka–{\L}ojasiewicz (KL) inequality at a point $\bar{\uvx} \in \operatorname{dom}(\partial \chi) $ if there exist a $\eta>0$, a neighborhood $\mathcal U$ of $\bar{\uvx}$, and a continuous concave function $ \varphi : [0, \eta) \to \mathbb{R}_+$ that is continuously differentiable on $(0,\eta)$ and satisfies  $\varphi(0) = 0$ and $ \varphi'(s) > 0$ for all $s \in (0, \eta)$, such that 
$$
\varphi'\big(|\chi(\uvx) - \chi(\bar{\uvx})|\big) \cdot \dist{\bm 0, \partial \chi(\uvx)} \geq 1.
$$
holds for all $\uvx \in \mathcal U \cap \{\uvx \in \mathbb{C}^N: |\chi(\uvx)-\chi(\bar{\uvx})|<\eta\}$.
Here, $\partial \chi(\uvx)$ denotes the subgradient of $\chi(\uvx)$,
and $\dist{\cdot,\cdot}$ denotes Euclidean distance. 
\end{definition}
In~\Cref{def:KLIneq}, $\varphi(s)$ is called the desingularization function.
If $\varphi(s)$ holds the form $\varphi(s)=cs^{1-t}$ for $t\in[0,1)$ and $c>0$,
then we say that $\chi(\uvx)$ has the KL property at $\bar{\uvx}$ with an exponent of $t$. 
\begin{assume}[$L$-Smooth $f$]
\label{assum:LipGrad}
Assume that $f: \mathbb{C}^n \to (-\infty, +\infty]$ is a proper, lower semicontinuous, and lower bounded function. Further assume that the gradient of $f$ is $L$-Lipschitz continuous. That is, $\forall \uvx_1,\,\uvx_2\in\mathbb C^N$, there exists a $L>0$ such that the following inequality holds:
	\begin{equation}
	\label{eq:gradLip:l}
		\|\nabla f(\uvx_1)-\nabla f(\uvx_2)\|\leq L \, \|\uvx_1-\uvx_2\|.
	\end{equation}
\end{assume}

\begin{lemma}[Majorizer of $f$ {\cite[Lemma 1]{hong2025CQNPMCSMRI}}]
\label{lemma:DescentLemma}
    Let $f:\,\mathbb C^N\rightarrow (-\infty,\infty]$ be an $L$-smooth function. Then for any $\uvx_1,
\uvx_2\in\mathbb C^N$, we have
\begin{equation}
    \label{eq:DescentLemma}
    f(\uvx_2)\leq f(\uvx_1)+\Re\big\{\langle \nabla f(\uvx_1),\uvx_2-\uvx_1\rangle\big\}+\frac{L}{2}\|\uvx_1-\uvx_2\|_2^2.
\end{equation}
\end{lemma}

\begin{lemma}[Bounded Hessian {\cite[Lemma 4]{hong2025CQNPMCSMRI}}]
	\label{lemma:boundedHessian}
    The approximate Hessian matrices $\umB_k$ generated by \Cref{alg:ModfiedZero:SR1} satisfy the following inequality
	$$\underline{\eta}\,\umI \preceq \umB_k \preceq \overline{\eta}\,\umI,$$
    where $0<\underline{\eta}<\overline{\eta}<\infty$.
\end{lemma}

\begin{lemma}	
\label{lemma:Ineq}
	By running~\Cref{alg:NGKS} for solving~\eqref{eq:CSMRIReco:rewrite}, we have the following inequality at $k$th iteration,
	\begin{equation}
		\label{eq:lemma:OptimalIneq}
	\begin{aligned}
	\Re\big\{\langle \nabla f(\uvx_k),\uvx_{k+1}-\uvx_k\rangle \big\}\leq&\,h(\uvx_k)-h(\uvx_{k+1})\\
	&-\frac{1}{2}\|\uvx_k-\uvx_{k+1}\|^2_{(\frac{2}{\stepsize_k}\umB_k-\underline{\eta}\umI_N)}.
\end{aligned}
	\end{equation}
	    \end{lemma}
\begin{lemma}
	\label{lemma:ineqs:ConvRate}
	Suppose the elements in $\{\phi_k>0\}_{k\geq 1}$ satisfy 
	$$
	\phi_{k+1}^{2t}\leq \gamma (\phi_k-\phi_{k+1})~\text{and}~\phi_{k+1}\leq \phi_k,
	$$
	where $t\in(0,1)$ and $\gamma>0$. Then we have  the following upper bounds for $\phi_{k+1}$,
	\begin{equation}
		\label{eq:UpperBoundPhi_k}
	\phi_{k+1}\leq 
	\begin{cases}
    \left(1-\frac{\phi_1^{2t-1}}{\gamma+\phi_1^{2t-1}}\right)^{k}\phi_1,&t\in(0,\frac{1}{2}] \\[5pt]
	\left(\phi_1^{1-2t}+\frac{(2t-1)(1-\sigma)^{2t}}{2\gamma}k\right)_+^{\frac{1}{1-2t}},& t\in(\frac{1}{2},1),
	\end{cases}
	\end{equation}
	where $\sigma \in (0,1)$ and $(\cdot)_+=\max(\cdot,0)$.
	\end{lemma}
\Cref{lemma:DescentLemma,lemma:boundedHessian} were already demonstrated in~\cite{hong2025CQNPMCSMRI}, so we omit their proofs here. The proofs of~\Cref{lemma:Ineq,lemma:ineqs:ConvRate} are provided in~\Cref{app:proof:lemma:Ineq,app:proof:lemma:ineqs:ConvRate}.
\Cref{lemma:ineqs:ConvRate} is used to establish the convergence rates of the cost function sequence for different values of $t$ in the KL inequality when running~\Cref{alg:NGKS}. \Cref{them:ConvResults:ksmallK,them:ConvResults:klargeK} summarize our
main convergence results.
\begin{theorem}[Descent properties of~\Cref{alg:NGKS}, $K\leq +\infty$]
\label{them:ConvResults:ksmallK}
Let $\stepsize_k\in(0,\frac{2\underline{\eta}}{\underline{\eta}+L})$
and $\Delta_k \defequ \min_{k'\leq k} \|\uvx_{k'+1}-\uvx_{k'}\|_2^2$.
Under~\Cref{assum:LipGrad}, by running $k<K$ iterations of~\Cref{alg:NGKS} 
to solve~\eqref{eq:CSMRIReco:rewrite},
we have 
\begin{itemize}
\item $
\Delta_k \leq \frac{F(\uvx_1)-F^*}{\upsilon k}~\text{and}~F(\uvx_{k+1})\leq F(\uvx_k),
$ where $\upsilon=\min_{k} \{{\underline{\eta}}/{\stepsize_{k}} -{(\underline{\eta}+L)}/{2}\}$, $F^*$ denotes the minimum of~\eqref{eq:CSMRIReco:rewrite}, and $\uvx_1$ is the initial iterate.
\item  $\|\uvx_{k+1}-\uvx_k\|\rightarrow 0$ as $k\rightarrow \infty$.
\end{itemize}
\end{theorem}

\begin{theorem}[Convergence rates, $K<+\infty$]
    \label{them:ConvResults:klargeK}
    Let $\stepsize_k\in(0, \frac{2\underline{\eta}}{\underline{\eta} + L})$ and $\mathcal{B}(\uvx^*, \Lambda) = \{ \uvx\in\mathbb C^N \mid \|\uvx - \uvx^*\| \leq \Lambda \}$. Under~\Cref{assum:LipGrad}, by running~\Cref{alg:NGKS} $k>K$ iterations
to solve~\eqref{eq:CSMRIReco:rewrite},
we have 
\begin{itemize}
  \item $\|\uvx_{k+1}-\uvx_k\|\rightarrow 0$ as $k\rightarrow \infty$ and all cluster points of the sequence $\{\uvx_k\}_{k> K}$ are critical points of~\eqref{eq:CSMRIReco:rewrite}.
    \item Assume $\uvx_k$ converges to $\uvx^*$ and $F_{\mathcal C}$ satisfies the KL inequality. There exists $\Lambda > 0$ such that $F_{\mathcal C}$ at $\bar \uvx = \uvx^*$ in a neighborhood of $\mathcal U$ containing $\mathcal B(\uvx^*, \Lambda)$. Then there also exists $K'>K$, such that, $\uvx_k \in \mathcal B(\uvx^*, \Lambda)$ and $| F_{\mathcal C}(\uvx_k) - F^* | < \eta$ for all $k \geq K'$. For $k\geq K'$, we have the following convergence rates of the cost function values for $t\in[0,1)$:
\begin{enumerate}
	\item $F(\uvx_{k+1})-F^*\leq \left (F_{K'}-\frac{1}{\gamma}(k-K'+1)\right)_+,$ $t=0$,
	\item $F(\uvx_{k+1})-F^*\leq \left(1-\frac{F_{K'}^{2t-1}}{F_{K'}^{2t-1}+\gamma}\right)^{k-K'+1}F_{K'},
	$\\ $t\in(0,\frac{1}{2}]$,
	\item $F(\uvx_{k+1})-F^* \leq  \left(F_{K'}^{1-2t}+q(k-K'+1)\right)_+^{\frac{1}{1-2t}},$ \\
	$t\in(\frac{1}{2},1)$.
\end{enumerate}
where $F_{K'}=F(\uvx_{K'})-F^*$, $\sigma\in(0,1)$, $\gamma=\max_{k}\left  ([c(L\stepsize_{k}+\overline{\eta})]^2)/(\upsilon(1-t)^2\stepsize_{k}^2)\right)$, and $q=(t-1/2)(1-\sigma)^{2t}/\gamma$.
\end{itemize}
\end{theorem}
\Cref{them:ConvResults:ksmallK} states that \PropM leads to $\|\uvx_{k+1}-\uvx_k\|\rightarrow 0$ as $k\rightarrow \infty$ for any $K$, either finite or infinite. Our first result in~\Cref{them:ConvResults:klargeK} establishes all cluster points of the sequence $\{\uvx_k\}_{k>K}$ are critical points of~\eqref{eq:CSMRIReco:rewrite} for finite $K$. The second result in~\Cref{them:ConvResults:klargeK} provides convergence rates of the cost values for  $t\in[0,1)$ after $K'>K$ iterations. During the algorithm's execution, all iterates remain in $\mathcal{C}$, so that $ F_{\mathcal{C}}(\uvx_k) = F(\uvx_k)$ for all $k$. For ease of notation, we use $F(\uvx)$ instead of $F_{\mathcal{C}}(\uvx)$ to express the convergence rates in~\Cref{them:ConvResults:klargeK}. Note that if $F(\uvx_{k+1})= F^*$, the left hand side is identically zero and therefore the convergence rate bounds in~\Cref{them:ConvResults:klargeK} hold trivially. 



Note that~\Cref{alg:NGKS} reduces to CQNPM~\cite{hong2025CQNPMCSMRI} when $K$ is finite and $k > K$. \Cref{them:ConvResults:klargeK} then extends the theoretical analysis in~\cite{hong2025CQNPMCSMRI} to a more general class of functions $F_C$. Specifically, our analysis relies on the KL inequality, which is weaker than the Polyak-{\L}ojasiewicz (PL) inequality\footnote{The PL inequality corresponds to a special case of the KL inequality with $t = \frac{1}{2}$.} used in~\cite{hong2025CQNPMCSMRI}.
\Cref{sec:NumericalExps:Conv} empirically studies the convergence behavior of \Cref{alg:NGKS}
to validate our theoretical analysis.


\section{Numerical Experiments} 
\label{sec:NumericalExps}

\begin{figure*}[t]
	\centering
	\includegraphics[width=0.8\textwidth]{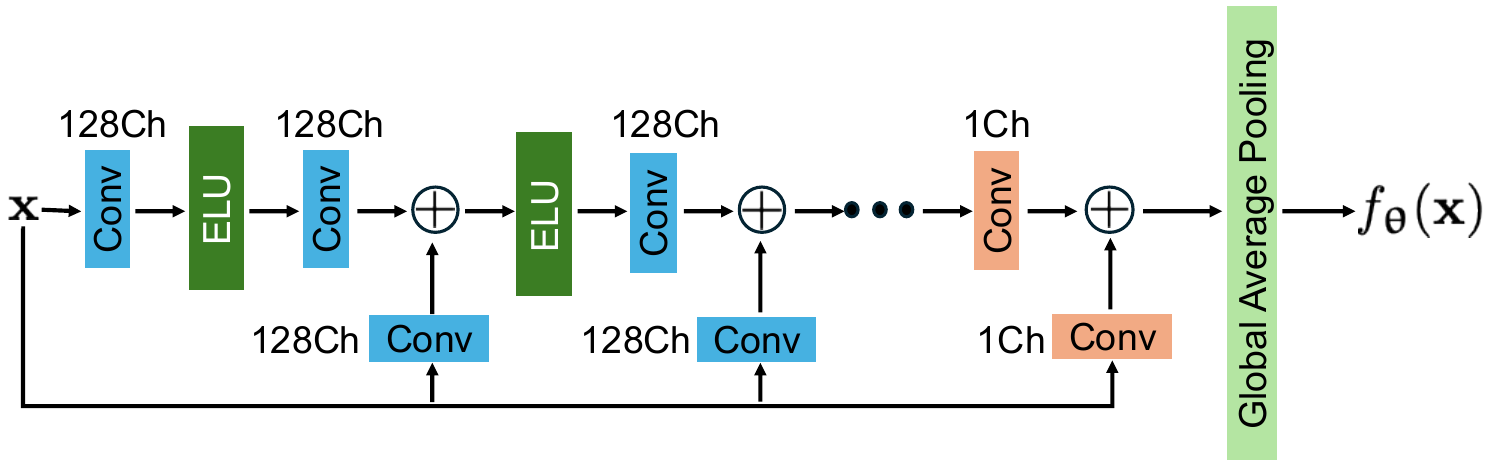}
	\caption{\MRcb{The neural network architecture used to construct the energy function 
$f_{\bm\uptheta}(\uvx)$ is based on \cite{cohen2021has}.
The convolutional kernels have size $3 \times 3$ with a stride of one.}}
	\label{fig:GDNNStructure}
\end{figure*}

\begin{figure*}[t]
    \centering
    \subfigure[1]{\includegraphics[width=0.16\textwidth]{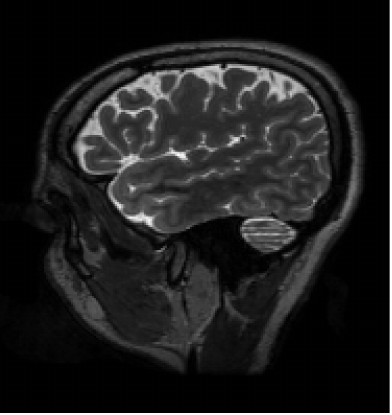}}
        \subfigure[2]{\includegraphics[width=0.16\textwidth]{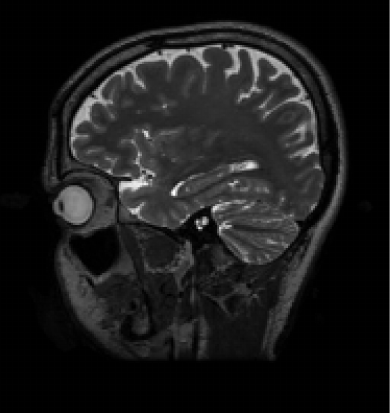}}
    \subfigure[3]{\includegraphics[width=0.16\textwidth]{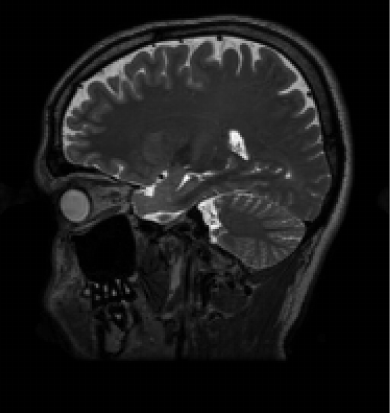}}
    \subfigure[4]{\includegraphics[width=0.16\textwidth]{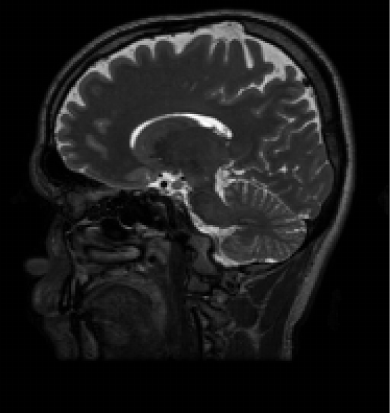}}
    \subfigure[5]{\includegraphics[width=0.16\textwidth]{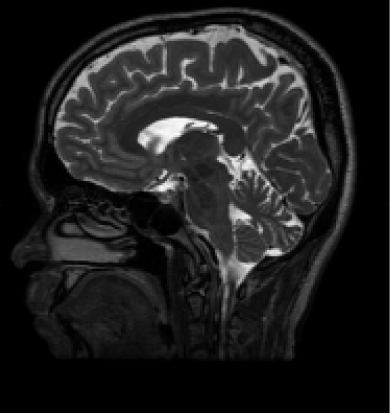}}
     \subfigure[6]{\includegraphics[width=0.16\textwidth]{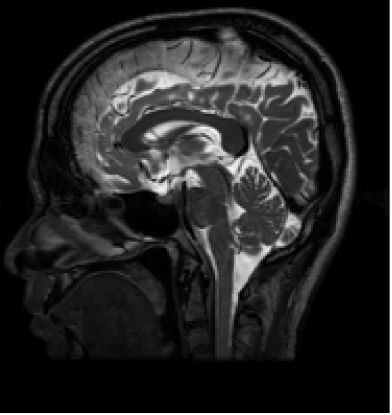}}

    \subfigure[1]{\includegraphics[width=0.16\textwidth]{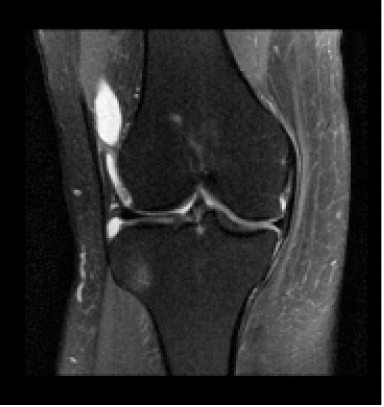}}
    \subfigure[2]{\includegraphics[width=0.16\textwidth]{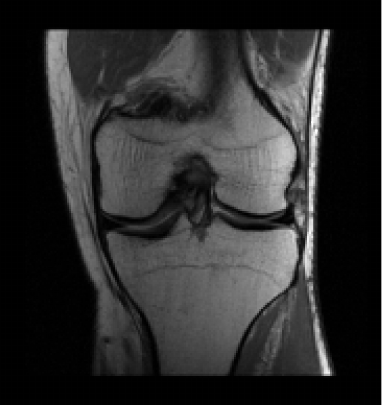}}
    \subfigure[3]{\includegraphics[width=0.16\textwidth]{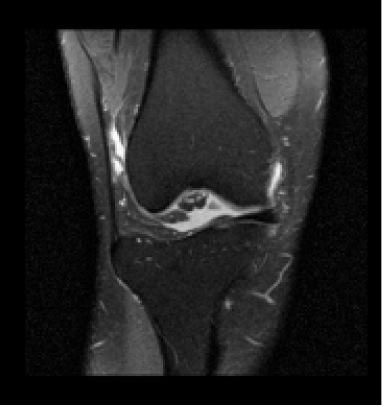}}
    \subfigure[4]{\includegraphics[width=0.16\textwidth]{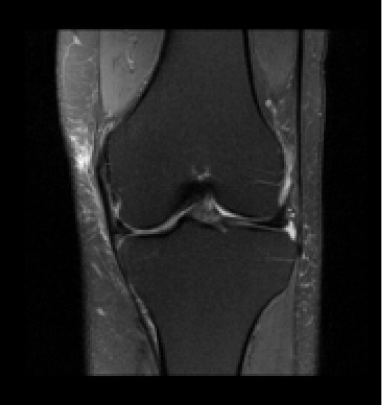}}
    \subfigure[5]{\includegraphics[width=0.16\textwidth]{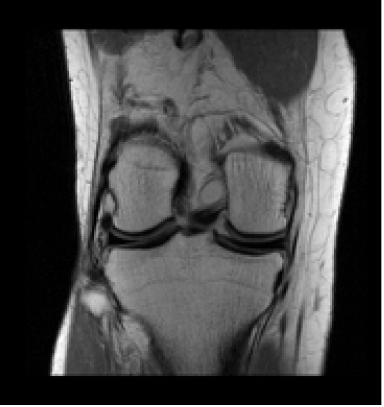}}
    \subfigure[6]{\includegraphics[width=0.16\textwidth]{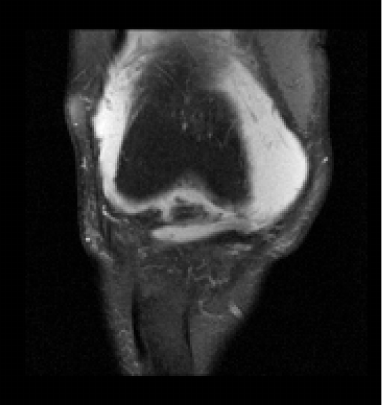}}
    \caption{The magnitude of the six brain and knee complex-valued ground truth images.}
    \label{fig:BrainKneeGT}
\end{figure*}

\begin{figure}[t]
\centering
\subfigure[Spiral]{
	\includegraphics[width=0.2\textwidth]{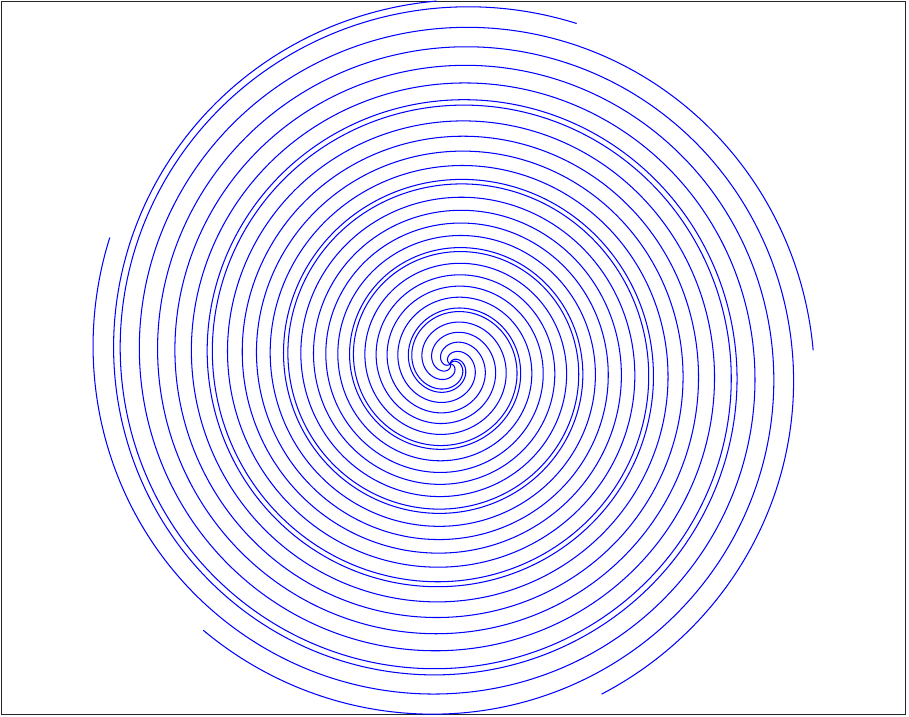}}
	\subfigure[Radial]{
	\includegraphics[width=0.2\textwidth]{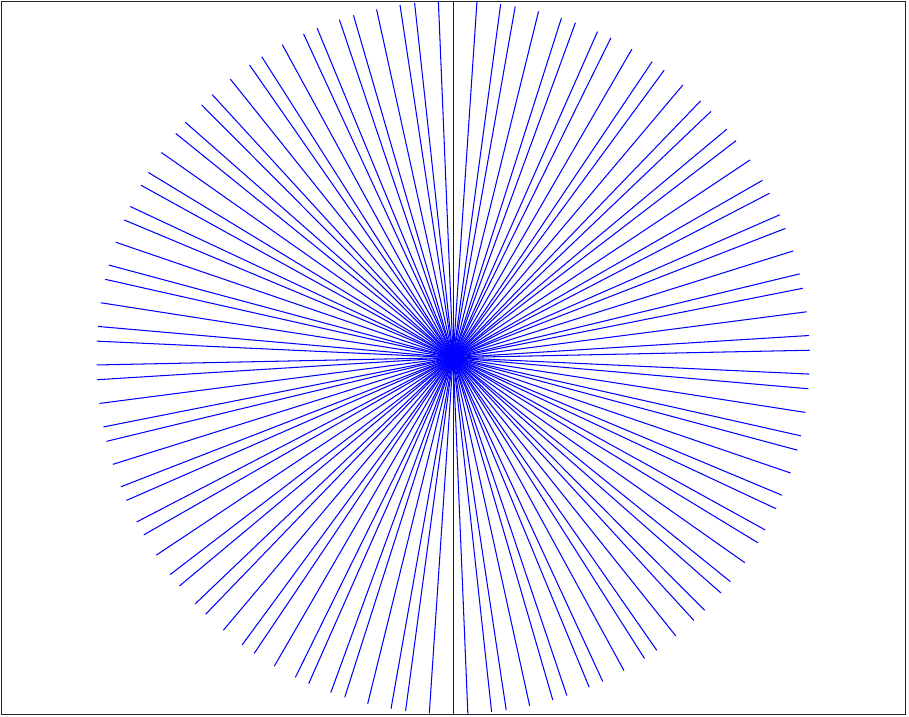}}
	\caption{The spiral (a) and radial (b) sampling trajectories.}
	\label{fig:TrjsMask}
\end{figure}

This section studies the performance of \PropM
for CS MRI reconstruction with spiral and radial sampling trajectories. Note that the experimental and algorithmic settings used here are similar to those in our previous work~\cite{hong2025CQNPMCSMRI}. For convenience, we briefly re-describe them in this paper. Then, we present the reconstruction results
and study the convergence behavior of \PropM empirically.

{\noindent \bf  Experimental Settings:} The performance of \PropM are evaluated on both brain and knee MRI datasets. For the brain images, we used the dataset from~\cite{aggarwal2018modl}, which contains 360 training and 164 test images. For the knee images, we adopted the multi-coil knee dataset from the NYU fastMRI~\cite{zbontar2018fastmri}. The ESPIRiT algorithm~\cite{uecker2014espirit} was used to obtain the complex-valued images from the raw k-space data. All images were then resized to a uniform \MRcb{image size} of $256 \times 256$ and normalized such that the maximum magnitude was one. The network architecture proposed in~\cite{cohen2021has} with the addition of bias terms is used to construct $f(\uvx)$, \MRcb{see \Cref{fig:GDNNStructure}. The number of layers is set to six. Note that other neural network architectures can also be used to train the gradient-driven denoiser. The only requirement is that the nonlinear activation functions be differentiable. For example, Hurault et al. \cite{hurault2021gradient} used DRUNet by replacing the ReLU activations with ELU to construct $f(\uvx)$.}  The noisy images were generated by adding i.i.d. Gaussian noise with variance $1/255$ to the clean images. \MRcb{The network was trained using the mean squared error loss.} We employed the ADAM optimizer~\cite{kingma2014adam} with an initial learning rate of $10^{-3}$, which was halved every $4,000$ iterations. Training was performed for a total of $18,000$ iterations with a batch size of $64$. Although we trained separate denoisers for the brain and knee datasets, different sampling trajectories used the same denoiser.

To assess reconstruction quality, we selected six brain and knee test images as ground truth. \Cref{fig:BrainKneeGT} displays the magnitudes of these images. For spiral acquisition, we used six interleaves with $1688$ readout points and $32$ coils. Radial acquisition employed $55$ spokes with golden-angle rotation, $1024$ readout points, and $32$ coils. \Cref{fig:TrjsMask} describes these sampling trajectories. To simulate k-space data, we applied the forward model to the ground-truth images and then added the complex i.i.d. Gaussian noise (zero mean, variance $10^{-4}$), resulting in an input SNR of approximately $21$\,dB. In the reconstruction, we employed coil compression~\cite{zhang2013coil} to reduce the number of coils from $32$ to $20$ virtual coils, thereby lowering the computational cost.
\MRcb{All experiments used simulated k-space data}
and were implemented in PyTorch and ran on an NVIDIA A100 GPU.

{\noindent \bf Algorithmic Settings:}
The previous work~\cite{hong2024complex} already showed that the accelerated proximal gradient method (dubbed APG)~\cite{li2015accelerated} is faster than the projected gradient descent method~\cite{cohen2021has} and the proximal gradient method~\cite{hurault2021gradient} for addressing~\eqref{eq:CSMRIReco:rewrite}.
Thus, we mainly compared \PropM with CQNPM~\cite{hong2024complex} and APG in this paper.
To test the applicability of \PropM for a constrained problem, we used the constraint set $\mathcal{C} = \{\uvx \mid \|\uvx\|_\infty \leq 1\}$ in all competing methods. However, in practical CS MRI reconstruction, we generally do not impose such a constraint.  For plots involving $F^*$, we ran APG for $500$ iterations and defined $F^* = F(\uvx_{500}) - \varepsilon$ for a small constant $\varepsilon > 0$.
Unless otherwise specified, we set $K=\texttt{Max\_Iter}$ in the following experiments. \Cref{alg:ModfiedZero:SR1} used $\delta = 10^{-8}$, $\nu_1=2\times 10^{-6}$, and $\nu_2=200$.

\subsection{Spiral Acquisition Reconstruction}
\label{sec:NumericalExps:Spiral}
\input{figs/SpiralBrain1CostPSNR}

\input{figs/SpiralBrain1RecoIm}

\begin{table*}[t]
    \centering
    \caption{PSNR performance of each method for reconstructing five additional brain test images with spiral acquisition.
    For APG, we report the maximum PSNR (within $150$ iterations), the corresponding number of iterations, and  wall time.
    For other methods, we report the earliest iteration count that exceeds the APG PSNR (along with its PSNR and wall time) in the first row. The PSNR and wall time at the $150$th iteration were summarized in the second row.
    Bold indicates the shortest wall time at which the PSNR of APG was exceeded. The PSNR values and wall time at the $150$th iteration of CQNPM and \PropM are marked with an underline. The {\color{blue}blue} digits denote the shortest wall time at the $150$th iteration.}
    \setlength{\tabcolsep}{4pt}
    \renewcommand{\arraystretch}{1.2}
    \begin{tabular}{p{2.1cm} @{} rrrr rrrr rrrr rrrr rrrr @{}}
        \toprule
  \multirow{2}*{\diagbox[innerwidth=1.83cm]{Methods}{Index}}
        & \multicolumn{3}{c}{ 2} & \multicolumn{3}{c}{ 3} & \multicolumn{3}{c}{ 4} 
        & \multicolumn{3}{c}{ 5} & \multicolumn{3}{c}{ 6} \\
        & PSNR$\uparrow$ & iter.$\downarrow$ & sec.$\downarrow$ 
        & PSNR$\uparrow$ & iter.$\downarrow$ & sec.$\downarrow$
        & PSNR$\uparrow$ & iter.$\downarrow$ & sec.$\downarrow$
        & PSNR$\uparrow$ & iter.$\downarrow$ & sec.$\downarrow$
        & PSNR$\uparrow$ & iter.$\downarrow$ & sec.$\downarrow$ \\
        \midrule
        APG &  $42.1$ & $150$ &  $86.3$ & 	$42.8$ & $150$ &  $86.4$	 &	$43.3$ &$150$  & 	$87.1$ &$42.0$	 & $150$ &   $86.4$  &$40.5$& $150$ &   	$88.7$ \\
        \midrule
        \multirow{2}{*}{CQNPM}      
        & $42.1$&$29$ & $16.9$ &$42.8$ &$30$ & $19.7$	& $43.3$	&$31$ &$19.1$  & $42.1$&$30$ & $17.6$ &	$40.5$ &$29$ &$17.6$ \\
       
 &$\underline{44.2}$	&$150$& $\underline{82.0}$ & $\underline{44.7}$ & $150$ &	$\underline{70.1}$ & $\underline{45.0}$ &$150$&	$\underline{65.5}$&	$\underline{44.0}$&$150$&  	 $\underline{68.9}$	&	$\underline{42.7}$&$150$& $\underline{77.0}$ \\
        \midrule
        \multirow{2}{*}{\PropM}     
        &$42.2$ &$49$& $\bm{2.4}$& $42.9$ &$50$&	 $\bm{2.4}$& $43.3$  &$50$& 	$\bm{2.4}$& $42.1$ &$51$&	 $\bm{2.5}$ & $40.5$ &$49$&$\bm{2.3}$\\
          &	$\underline{44.2}$ &$150$&$\underline{\color{blue}8.4}$&	$\underline{44.7}$	&$150$&$\underline{\color{blue}8.2}$&	$\underline{45.1}$&$150$&$\underline{\color{blue}8.2}$	&		$\underline{43.9}$&$150$&$\underline{\color{blue}8.3}$	&		$\underline{42.5}$&$150$&$\underline{\color{blue}8.2}$\\
        \bottomrule
    \end{tabular}
    \label{tab:SpiralBrainOthers}
\end{table*}

\Cref{fig:SpiralBrain1CostPSNR} summarizes the cost and PSNR values versus the number of iterations and wall time for each method on the brain $1$ image. \Cref{fig:SpiralBrain1CostPSNR}.s (a) and (c) show CQNPM was the fastest algorithm in terms of the number of iterations.
\PropM achieved similar results to CQNPM after enough iterations.
\Cref{fig:SpiralBrain1CostPSNR}s (b) and (d) report the cost and PSNR values versus wall time, where \PropM was the fastest algorithm in terms of wall time.
In multi-coil CS MRI reconstruction with non-Cartesian sampling, computing $\umA\uvx$ and $\umA^\HTrans\uvx$ typically dominate the computational cost.
CQNPM (respectively, APG) requires solving a weighted proximal mapping (respectively, a proximal mapping) at each iteration, which involves multiple evaluations of $\umA\uvx$ and $\umA^\HTrans\uvx$. In contrast, \PropM requires only a single evaluation of $\umA\uvx$ and $\umA^\HTrans\uvx$ per iteration, which significantly reduces computational cost while maintaining relatively fast convergence in terms of the number of iterations.


\Cref{fig:SpiralBrain1:visual} presents the reconstructed images
and the corresponding error maps at the $50$th and $100$th iterations of each method.
\PropM eventually achieved a reconstruction quality similar to that of CQNPM with same number of iterations,
but with significantly less wall time.
\Cref{tab:SpiralBrainOthers} summarizes the PSNR performance on the rest of five brain images within $150$ iterations. Clearly, we observed that \PropM was approximately $7\times$ faster than CQNPM in terms of wall time required to exceed the performance of APG within $150$ iterations. Moreover, \PropM achieves nearly the same performance as CQNPM at the $150$th iteration, while requiring approximately $9\times$ less time, illustrating the superior performance of our method. The supplementary material includes additional results on the knee images with spiral acquisition, \MRcb{as well as the structural similarity index measure (SSIM) metrics}, which exhibit similar behavior. 

\subsection{Radial Acquisition Reconstruction}
\label{sec:NumericalExps:Radial}

\input{figs/RadialKnee7CostPSNR}
\input{figs/RadialKnee7RecoIm}

\begin{table*}[t]
    \centering
    \caption{PSNR performance of each method for reconstructing five additional knee test images with radial acquisition.
    For APG, we report the maximum PSNR (within $100$ iterations), the corresponding number of iterations, and wall time.
    For other methods, we report the earliest iteration that exceeds the APG PSNR (along with its PSNR and wall time) in the first row. 
     The PSNR and wall time at the $100$th iteration were summarized in the second row. Bold indicates the shortest wall time at which the PSNR of APG was exceeded. The PSNR values and wall time at the $100$th iteration of CQNPM and \PropM are marked with an underline. The {\color{blue}blue} digits denote the shortest wall time at the $100$th iteration.}
    \setlength{\tabcolsep}{4pt}
    \renewcommand{\arraystretch}{1.2}
    \begin{tabular}{p{2.1cm} @{} rrrr rrrr rrrr rrrr rrrr @{}}
        \toprule
  \multirow{2}*{\diagbox[innerwidth=1.83cm]{Methods}{Index}}
        & \multicolumn{3}{c}{ 2} & \multicolumn{3}{c}{ 3} & \multicolumn{3}{c}{ 4} 
        & \multicolumn{3}{c}{ 5} & \multicolumn{3}{c}{ 6} \\
        & PSNR$\uparrow$ & iter.$\downarrow$ & sec.$\downarrow$ 
        & PSNR$\uparrow$ & iter.$\downarrow$ & sec.$\downarrow$
        & PSNR$\uparrow$ & iter.$\downarrow$ & sec.$\downarrow$
        & PSNR$\uparrow$ & iter.$\downarrow$ & sec.$\downarrow$
        & PSNR$\uparrow$ & iter.$\downarrow$ & sec.$\downarrow$ \\
        \midrule
        APG       
        &$42.6$	&$100$&$61.0$ &  $41.4$ &$100$&	$61.3$ 	&	$44.6$ &$100$&$60.7$ & $42.1$ &$100$&	$63.6$ &$44.2$ &$100$& $64.7$\\
        \midrule
        \multirow{2}{*}{CQNPM}     
        &$42.7$  &$26$&  $15.8$	&  $41.4$ &$30$&   $19.1$ & $44.7$ &$25$& $16.4$ & 	$42.2$&$28$&	$17.0$&  	 $44.2$&$25$&	$16.3$\\
        &$\underline{44.1}$ &$100$&$\underline{59.9}$& $\underline{41.9}$	&$100$& $\underline{64.3}$&$\underline{45.5}$ &$100$&$\underline{65.4}$&	$\underline{44.1}$	&$100$&$\underline{61.9}$	 &$\underline{44.9}$ &$100$& $\underline{66.3}$\\
        \midrule
        \multirow{2}{*}{\PropM}
              &$42.7$ &$41$&$\bm{2.0}$& 	$41.4$ &$33$&	$\bm{1.6}$ & $44.7$ &$38$& $\bm{1.8}$	&	$42.3$ &$36$&	 $\bm{1.8}$ & 	$44.2$ &$39$&$\bm{1.9}$\\
              &$\underline{44.1}$ &$100$&  $\underline{\color{blue}5.6}$&	$\underline{43.3}$	&$100$&$\underline{\color{blue}5.5}$&$\underline{45.5}$&$100$&$\underline{\color{blue}5.5}$&	$\underline{44.4}$	&$100$&$\underline{\color{blue}5.7}$&$\underline{44.9}$&$100$&$\underline{\color{blue}5.5}$\\
        \bottomrule
    \end{tabular}
    \label{tab:RadialKneeOthers}
\end{table*}

\Cref{fig:RadialKnee1CostPSNR} presents the cost and PSNR values versus the number of iterations and wall time of each method on the knee $1$ image with radial acquisition. \Cref{fig:RadialKnee1CostPSNR}s (a) and (c) show CQNPM converged faster than APG in terms of the number of iterations. In this experimental setting, we found that CQNPM made faster progress than \PropM in the early iterations. Then \PropM exceeded CQNPM in the later iterations. This observation is slightly different from \Cref{fig:SpiralBrain1CostPSNR}. Although all methods have convergence guarantees under the same assumptions, we cannot guarantee that their iterates follow the same trajectory. One possible explanation is that using a subspace in \PropM may sometimes act as an additional constraint, guiding the iterates along a more favorable path toward a minimizer. The detailed study of this direction is beyond the scope of this paper and is left for future work.  \Cref{fig:RadialKnee1CostPSNR}s (b) and (d) display the cost and PSNR values of each method versus  wall time. Evidently, \PropM converged faster than others in terms of  wall time, which is consistent with the previous observation. 
  
\Cref{fig:RadialKnee1:visual} reports the reconstructed images of each method at $50$th and $100$th iterations. From this experiment, we observed that \PropM demonstrated the best visual quality among all methods. \Cref{tab:RadialKneeOthers} reports the PSNR and wall time performance of each method on the remaining five knee test images. We ran APG for $100$ iterations and then compared how many iterations were required by the other methods to exceed the PSNR value achieved by APG. \Cref{tab:RadialKneeOthers} also reports the PSNR values and wall time of CQNPM and \PropM at $100$ iterations. Consistently, we observed similar beharior as in \Cref{tab:SpiralBrainOthers}.
The supplementary material includes additional results on the brain images with radial acquisition,
\MRcb{as well as the SSIM metrics} that show similar trends.

\subsection{Effect of \texorpdfstring{$K$}{TEXT} and Convergence Validation}
\label{sec:NumericalExps:Conv}

\input{figs/SpiralBrainConvVal}

This part empirically studies the effect of $K$ and the convergence behavior of \PropM using the brain $1$ image and spiral acquisition settings as in \Cref{fig:SpiralBrain1CostPSNR}. \Cref{fig:SpiralBrain1VarK} reports the PSNR values versus the number of iterations and  wall time for \PropM with varying $K$ and CQNPM. We observed that different values of $K$ converged to similar PSNR values as CQNPM.  \Cref{fig:SpiralBrain1VarK}(b) presents the PSNR values versus wall time, where we observed that larger values of $K$ led to faster convergence compared with smaller ones. This observation is consistent with our earlier results, as \PropM avoids solving a weighted proximal mapping for iterations $k \leq K$.

We have now empirically validated our theoretical analysis. \Cref{fig:ConvVal} presents the cost values and the values of $\Delta_k/\Delta_1$ for \PropM with spiral acquisition on six brain test images. As expected, the cost values converged to a constant across all test images, and $\Delta_k/\Delta_1 \rightarrow 0$, consistent with our theoretical analysis.

\section{Conclusion}
\label{sec:Conclusion}
A well-established theoretical foundation is especially important for ensuring reliability in medical imaging applications. Compared with the PnP and RED frameworks, gradient-driven denoisers offer a significantly stronger theoretical foundation. In particular, the only required assumptions are the differentiability of $f$  and the Lipschitz continuity of~$\nabla f$, which are easier to satisfy in practice.  To efficiently solve the associate nonconvex minimization problem, we developed a generalized Krylov subspace method with convergence guarantees in nonconvex settings. Numerical experiments on multi-coil CS-MRI reconstruction with non-Cartesian sampling trajectories demonstrate that the proposed method can recover images within seconds on a GPU platform. This significantly improves the efficiency of solving the associated optimization problem and enhances the practical applicability of gradient-driven denoisers. 


\appendices
\crefalias{section}{appendix}

\section{Proof of \texorpdfstring{\Cref{lemma:Ineq}}{Lemma 3}}
\label{app:proof:lemma:Ineq}
Since $\umB_k\succ 0$ (cf. \Cref{lemma:boundedHessian}), we know~the objective function in \eqref{eq:CSMRIReco:origP:GradDenoiser:MM} is $\underline{\eta}$-strongly convex. By combining with the fact that $\umV_k^\HTrans \umV_k=\umI_k$ and the $\underline{\eta}$-strongly convex inequality, we have the following inequality at $k$th iteration for $\forall\,\uvx =\umV_k\vbeta,\,\uvx\in\mathcal C$:  
\begin{equation}
\label{eq:OptimalCondStronglyConv:beta}
\begin{aligned}
	\Re\Big\{ \big\langle \umV_k^\HTrans \big[\nabla h(\uvx_{k+1})+\nabla \bar f(\uvx_{k+1},\uvx_k,\umB_k,\stepsize_k)&\\
	+\frac{\underline{\eta}}{2}(\umV_k\vbeta-\uvx_{k+1})\big], \vbeta-\vbeta_k \big \rangle \Big\}&\geq 0.
\end{aligned}	
\end{equation}
Letting $\vbeta = \bmat \vbeta_{k-1}\\0 \emat$,  we rewrite \eqref{eq:OptimalCondStronglyConv:beta} as
\begin{equation}
\label{eq:OptimalCondStronglyConv:x}
\begin{aligned}
	\Re\Big\{ \big\langle \big[\nabla h(\uvx_{k+1})+\frac{1}{\stepsize_k}\umB_k(\uvx_{k+1}-\uvx_k) +\nabla f(\uvx_k)&\\
	 +\frac{\underline{\eta}}{2}(\uvx_k-\uvx_{k+1})\big], \uvx_k-\uvx_{k+1} \big \rangle \Big\}&\geq 0.
\end{aligned}
\end{equation}
Note that if $\|\tilde{\uvr}_k\|= 0$, we choose $\vbeta = \vbeta_{k-1}$ and~\eqref{eq:OptimalCondStronglyConv:x} is still held. By reorganizing \eqref{eq:OptimalCondStronglyConv:x} and using the convexity of $h(\uvx)$ ($h(\uvx_k)\geq h(\uvx_{k+1})+\Re\{\langle \nabla h(\uvx_{k+1}),\uvx_k-\uvx_{k+1}\rangle\}$), we get the desired result
\begin{equation}
\label{eq:OptimalCondStronglyConv:inequaity}
\begin{aligned}
	\Re\big\{\langle \nabla f(\uvx_k),\uvx_{k+1}-\uvx_k\rangle \big\}\leq&\,h(\uvx_k)-h(\uvx_{k+1})\\
	& -\frac{1}{2}\|\uvx_k-\uvx_{k+1}\|^2_{(\frac{2}{\stepsize_k}\umB_k-\underline{\eta}\umI_N)}.
\end{aligned}	
\end{equation}

\section{Proof of \texorpdfstring{\Cref{lemma:ineqs:ConvRate}}{Lemma 4}}
\label{app:proof:lemma:ineqs:ConvRate}
Our goal is to derive an upper bound for $\phi_{k+1}$ by using the facts that $\phi_k-\phi_{k+1}\geq \phi_{k+1}^{2t}/\gamma$ and $0<\phi_{k+1}\leq \phi_k$. Rewrite $\phi_k-\phi_{k+1}\geq \phi_{k+1}^{2t}/\gamma$ as
$\phi_k\geq\phi_{k+1}(1+\frac{1}{\gamma}\phi_{k+1}^{2t-1})$. Considering $t\in(0,1/2)$, we know  $\phi_{k+1}^{2t-1}$ is monotonically decreasing since $2t-1<0$. So we have $\phi_{k+1}^{\,2t-1} \geq \phi_1^{2t-1},$ which implies
$
\phi_k \geq \phi_{k+1} \left( 1 + \frac{1}{\gamma}\phi_1^{2t-1} \right),
$
yielding
$
\phi_{k+1} \leq
\left( 1 - \frac{\phi_1^{\,2t-1}}{\gamma + \phi_1^{\,2t-1}} \right)^{k} \phi_1.
$ If $t = \frac{1}{2}$, we have 
$
\phi_{k+1} \leq \gamma (\phi_k - \phi_{k+1}),
$
which yields
$
\phi_{k+1} \leq \frac{\gamma}{1 + \gamma} \phi_k.
$
Therefore, we can establish the desired result immediately:
$
\phi_{k+1} \le \left( 1 - \frac{1}{1 + \gamma} \right)^k \phi_1.
$


Denote by $\psi(x)=x^{1-2t}$, where $x>0$. Let $\bar t=2t-1$. Using the mean value theorem, we have 
\begin{equation}
\label{eq:boundIneq:mean}
	\psi(\phi_{k+1})-\psi(\phi_k)= -\bar t\, {\bar\phi_k}^{-\bar t-1}(\phi_{k+1}-\phi_k),
\end{equation}
with $\phi_{k+1}\leq \bar\phi_k \leq \phi_k$. Since ${\bar\phi_k}^{-\bar t-1}$ is monotonically decreasing and $\phi_k-\phi_{k+1}\geq \phi_{k+1}^{2t}/\gamma$, we can get the following inequalities from \eqref{eq:boundIneq:mean} for $t\in(1/2,1)$
\begin{equation}
\label{eq:boundIneq:first}
\begin{array}{rcl}
	\psi(\phi_{k+1})-\psi(\phi_k)&\geq & \bar t\, \phi_k^{-2t}(\phi_k-\phi_{k+1})\\[5pt]
	&\geq& \phi_k^{-2t}\phi_{k+1}^{2t}\frac{\bar t}{\gamma}.
\end{array}
\end{equation}
Since $0<\phi_{k+1}\leq \phi_k$, we have $\phi_{k+1}/\phi_k\leq 1$. Suppose we run $k$ iterations. For any $\sigma\in(0,1)$, we can split the whole iterate indices into two subsets $\mathcal I_1$ and $\mathcal I_2$ such that $\mathcal I_1=\{k' \mid \phi_{k'+1}/\phi_{k'} \leq 1-\sigma\}$ and $\mathcal I_2=\{k' \mid \phi_{k'+1}/\phi_{k'} > 1-\sigma\}$. So, we know that either $|\mathcal I_1|\geq k/2$ or $|\mathcal I_2|\geq k/2$.

If  $|\mathcal I_1|\geq k/2$, we get 
\begin{equation}
\label{eq:BoundGeo}
	\phi_{k+1}\leq (1-\sigma)^{k/2}\phi_1.
\end{equation} 
Next, we consider $|\mathcal I_2|\geq k/2$. By summing up~\eqref{eq:boundIneq:first} from $1$ to $k$, we reach
 $$
\psi(\phi_{k+1})\geq \psi(\phi_1)+\frac{(2t-1)(1-\sigma)^{2t}}{2\gamma}k.
$$
Using the definition of $\psi(\cdot)$ and the fact that $1-2t<0$, we derive 
\begin{equation}
\label{eq:BoundLinear}
	\phi_{k+1}\leq \left(\phi_1^{1-2t}+\frac{(2t-1)(1-\sigma)^{2t}}{2\gamma}k\right)_+^{\frac{1}{1-2t}}.
\end{equation} 
If $k$ is large enough, the bound in \eqref{eq:BoundGeo} is smaller than that in \eqref{eq:BoundLinear}, yielding the desired result. 

\newpage
\section{Proof of \texorpdfstring{\Cref{them:ConvResults:ksmallK}}{Theorem 1}}
By using \Cref{lemma:DescentLemma}, we have the following inequalities 
\begin{equation}
\label{App:eq:DescentIneq}
\begin{aligned}
	f(\uvx_{k+1})\leq &\,f(\uvx_k)+\frac{L}{2}\|\uvx_{k+1}-\uvx_k\|_2^2  \\[3pt]
	&\,+\Re\{\langle \nabla f(\uvx_k),\uvx_{k+1}-\uvx_k\rangle\} \\
	\leq &\, f(\uvx_k)+h(\uvx_k)-h(\uvx_{k+1})\\
	&\,-\frac{1}{2}\|\uvx_k-\uvx_{k+1}\|^2_{\left(2\umB_k/\stepsize_k-(\underline{\eta}+L)\umI_N\right)}.
\end{aligned}	
\end{equation}
The second inequality comes from \Cref{lemma:Ineq}. Reorganizing~\eqref{App:eq:DescentIneq}, we get
$$
\frac{1}{2}\|\uvx_k-\uvx_{k+1}\|^2_{\left(2\umB_k/\stepsize_k-(\underline{\eta}+L)\umI_N\right)}\leq F(\uvx_k)-F(\uvx_{k+1}).
$$
Letting $\stepsize_k < \frac{2\underline{\eta}}{\underline{\eta}+L}$, $\upsilon=\min_{k} \{{\underline{\eta}}/{\stepsize_{k'}}-{(\underline{\eta}+L)}/{2}\}$, and using~\Cref{lemma:boundedHessian}, we reach 
\begin{equation}
	\label{eq:iter:cost:descent}
\upsilon \|\uvx_k-\uvx_{k+1}\|_2^2\leq F(\uvx_k)-F(\uvx_{k+1}).
\end{equation}
Since $\upsilon > 0$, we have $F(\uvx_{k+1})\leq F(\uvx_k)$. Summing up \eqref{eq:iter:cost:descent} from $k'=1$ to $k$, we get
\begin{equation}
\label{eq:iterdiff:summation}
\begin{aligned}
\sum_{k'=1}^k \upsilon \|\uvx_{k'}-\uvx_{k'+1}\|_2^2\leq& \, F(\uvx_1)-F(\uvx_{K+1})\\
\leq & \,F(\uvx_1)-F^*,
\end{aligned}	
\end{equation}
where $F^*$ denotes the minimal value of $F(\uvx)$. Letting $\Delta_k = \min_{k'\leq k} \{\|\uvx_{k'}-\uvx_{k'+1}\|_2^2\}$, we get the desired result 
\begin{equation}
\label{app:eq:Delta_k:ksmallK}
    \Delta_k \leq \frac{F(\uvx_1)-F^*}{\upsilon k}.
\end{equation}
Let $k\rightarrow \infty$, we get $\Delta_k\rightarrow 0$. Together with the summation in \eqref{eq:iterdiff:summation}, we obtain $\|\uvx_{k+1} - \uvx_k\| \rightarrow 0$ as $k \rightarrow \infty$.

\section{Proof of \texorpdfstring{\Cref{them:ConvResults:klargeK}}{Theorem 2}}
For $k > K$, we have $\umV_k = \umI_N$, so $\umV_k^H \umV_k = \umI_N$ still holds. Therefore,~\eqref{eq:iterdiff:summation} and~\eqref{app:eq:Delta_k:ksmallK} remain valid for $k > K$. Consequently, we still have $\|\uvx_{k+1} - \uvx_k\| \rightarrow 0$  as $k \rightarrow \infty$. Next, we prove that all cluster points of the sequence $\{\uvx_k\}_{k>K}$ are critical points of~\eqref{eq:CSMRIReco:rewrite}.

Let
$
G(\vbeta)=\left \| \bar{\umA}_k\vbeta - \bar{\uvy}_k\right \|_2^2
$
denote the cost function of~\eqref{eq:CSMRIReco:consP:GradDenoiser:MM:Beta} with $\bar{\umA}_k=\bmat\umA \\ \bar{\umB}_k^{\frac{1}{2}} \emat$ and $\bar{\uvy}_k=\bmat \uvy \\  \bar{\umB}_k^{\frac{1}{2}}\uvw_k \emat.$ Then we rewrite~\eqref{eq:CSMRIReco:consP:GradDenoiser:MM:Beta} as an unconstrained problem, i.e.,
\begin{equation}
\label{eq:CSMRIReco:UnconsP:GradDenoiser:MM:Beta}
	\vbeta_k=\arg\min_{\vbeta} G(\vbeta)+\iota_{\mathcal C}(\vbeta).
\end{equation}
From the first-order optimality condition of~\eqref{eq:CSMRIReco:UnconsP:GradDenoiser:MM:Beta} and using the fact that $\uvx_{k+1} = \vbeta_k$, we have
$$
\bm 0\in \nabla h(\uvx_{k+1}) +\partial \iota_{\mathcal C}(\uvx_{k+1}) + \nabla f(\uvx_k) + \frac{1}{\stepsize_k}\umB_k(\uvx_{k+1}-\uvx_k)
$$
which implies
\begin{equation}
\label{eq:FirstOrderIneqInCond}
\begin{aligned}
 \nabla f(\uvx_{k+1})- \nabla f(\uvx_k)&\\
\,\,+\frac{1}{\stepsize_k}\umB_k(\uvx_k-\uvx_{k+1})\in &\,  \nabla h(\uvx_{k+1}) +\nabla f(\uvx_{k+1})\\
&\,+\partial \iota_{\mathcal C}(\uvx_{k+1}).
\end{aligned}	
\end{equation}
Here, we use the definitions of $\bar f(\uvx,\uvx_k,\umB_k,\stepsize_k)$ and $h(\uvx)$.

Note that $F_{\mathcal C}(\uvx)=F(\uvx)+\iota_{\mathcal C}(\uvx)$ with $F(\uvx)=h(\uvx)+f(\uvx)$. By using~\eqref{eq:FirstOrderIneqInCond}, we have
\begin{equation}
\label{eq:gradIterVkIneq}
	\begin{aligned}
	\dist{\bm 0, \partial F_{\mathcal C}(\uvx_{k+1})}\leq &\big \|\nabla f(\uvx_{k+1})- \nabla f(\uvx_k)\\
	&\,+\frac{1}{\stepsize_k}\umB_k(\uvx_k-\uvx_{k+1})\big \|,\\
    	\leq &\,\big \|\nabla f(\uvx_{k+1})- \nabla f(\uvx_{k})\big\|
\\
	&\,+\frac{\overline{\eta}}{\stepsize_{k}}\|\uvx_{k}-\uvx_{k+1}\|\\
	\leq &\frac{L\stepsize_{k}+\overline{\eta}}{\stepsize_{k}}\|\uvx_{k}-\uvx_{k+1}\|.
\end{aligned}
\end{equation}
Notice that $\|\uvx_{k}-\uvx_{k+1}\|\rightarrow 0$ for $k\rightarrow \infty$ and that $\frac{L\stepsize_{k}+\overline{\eta}}{\stepsize_{k}}$ remains finite. So we have $\dist{\bm 0,\partial F_{\mathcal C}(\uvx_{k+1})}\rightarrow 0$ for $k\rightarrow \infty$, which implies that all cluster points of $\{\uvx_k\}_{k>K}$ are critical points of~\eqref{eq:CSMRIReco:rewrite}. This completes the proof of the first term.

Since $F_{\mathcal C}$ satisfies the KL inequality and $\uvx_k$ is converging to $\uvx^*$, there exist $K'>K$ and $\Lambda>0$ such that, for all $k \geq K'$, we have $\uvx_k\in\mathcal{B}(\uvx^*, \Lambda)$, where $\mathcal{B}(\uvx^*, \Lambda) = \{ \uvx\in\mathbb C^N \mid \|\uvx - \uvx^*\| \leq \Lambda \}$, and $| F_C(\uvx_k) - F^* | < \eta$. By letting $\bar{\uvx}=\uvx^*$, $\varphi(s)=cs^{1-t}$, and  the assumption $\mathcal{B}(\uvx^*, \Lambda)\subseteq \mathcal U$, we get
\begin{equation}
\label{eq:KLIneqTrueF}
\begin{aligned}
	(F_{\mathcal C}(\uvx_{k+1})-F^*)^{2t} \leq &c^2(1-t)^2\dist{\bm 0,\partial F_{\mathcal C}(\uvx_{k+1})}^2\\
	\overset{\eqref{eq:gradIterVkIneq}}{\leq} & \frac{[c(L\stepsize_{k}+\overline{\eta})]^2}{(1-t)^2\stepsize_{k}^2}\|\uvx_{k}-\uvx_{k+1}\|_2^2\\
	\overset{\eqref{eq:iter:cost:descent}}{\leq}
 & \frac{[c(L\stepsize_{k}+\overline{\eta})]^2}{\upsilon(1-t)^2\stepsize_{k}^2}(F(\uvx_{k})-F(\uvx_{k+1})).
\end{aligned}	
\end{equation}
Note that during the algorithm's progress, all iterates remain in $\mathcal{C}$, so that $F_{\mathcal{C}}(\uvx) = F(\uvx)$. For simplicity, we write $F(\uvx)$ instead of $ F_{\mathcal{C}}(\uvx)$ in what follows. Denote by $\gamma = \max_{k}\left  ([c(L\stepsize_{k}+\overline{\eta})]^2)/(\upsilon(1-t)^2\stepsize_{k}^2)\right)$. For $t=0$, we have 
$$
F(\uvx_{k+1})-F^*\leq F(\uvx_{k})-F^*-\frac{1}{\gamma}, 
$$
resulting in
$$
F(\uvx_{k+1})-F^*\leq \left (F(\uvx_{K'})-F^*-\frac{1}{\gamma}(k-K'+1)\right)_+. 
$$
By using~\Cref{lemma:ineqs:ConvRate}, we get the desired results for $t\in(0,1)$. 


\bibliographystyle{IEEEtran}
\bibliography{Refs}

\end{document}

%% file: defs.tex
\usepackage{subfigure}

\usepackage{graphicx}
\usepackage{url}
\usepackage{bm}
\usepackage[comma,numbers,square,sort&compress]{natbib}
\usepackage{hyperref} 
\usepackage{amsmath,amssymb,amsthm}  
\usepackage{paralist}

\usepackage{algorithm}
\usepackage{algorithmic}      
\usepackage{multirow}
\renewcommand{\algorithmicrequire}{\textbf{Initialization:}}
\renewcommand{\algorithmicensure}{\textbf{Iteration:}}
\renewcommand{\algorithmiclastcon}{\textbf{Output:}} 
\usepackage{upgreek}

\usepackage{cleveref}
\Crefname{figure}{Fig.}{Figs.}

\crefname{assume}{assumption}{assumptions}

\newtheorem{lemma}{Lemma}

\newtheorem{theorem}{Theorem}
\newtheorem{definition}{Definition}

\newtheorem{assume}{Assumption}

\newcommand{\e}{\begin{equation}}
\newcommand{\ee}{\end{equation}}
\newcommand{\en}{\begin{equation*}}
\newcommand{\een}{\end{equation*}}
\newcommand{\eqn}{\begin{eqnarray}}
\newcommand{\eeqn}{\end{eqnarray}}
\newcommand{\bmat}{\begin{bmatrix}}
\newcommand{\emat}{\end{bmatrix}}
\newcommand{\BIT}{\begin{itemize}}
\newcommand{\EIT}{\end{itemize}}

\newcommand{\Trans}{{\mathrm{T}}}  %
\newcommand{\HTrans}{{\mathrm{H}}}  %

\newcommand{\defequ}{\triangleq}

\newcommand{\blmath}[1]{\bm{\mathrm{#1}}}

\newcommand{\uvb}{\blmath{b}}

\newcommand{\uvm}{\blmath{m}}

\newcommand{\uvr}{\blmath{r}}
\newcommand{\uvs}{\blmath{s}}

\newcommand{\uvu}{\blmath{u}}
\newcommand{\uvv}{\blmath{v}}
\newcommand{\uvw}{\blmath{w}}
\newcommand{\uvx}{\blmath{x}}
\newcommand{\uvy}{\blmath{y}}
\newcommand{\uvz}{\blmath{z}}

\newcommand{\umA}{\blmath{A}}
\newcommand{\umB}{\blmath{B}}

\newcommand{\umF}{\blmath{F}}

\newcommand{\umI}{\blmath{I}}

\newcommand{\umP}{\blmath{P}}

\newcommand{\umS}{\blmath{S}}

\newcommand{\umV}{\blmath{V}}
\newcommand{\umW}{\blmath{W}}

\newcommand{\vbeta}{\bm\upbeta}

\newcounter{oursection}

\usepackage{xspace}

\newcommand{\dist}[1]{\operatorname{dist}\big(#1\big)}

%% file: figs/SpiralBrain1CostPSNR.tex
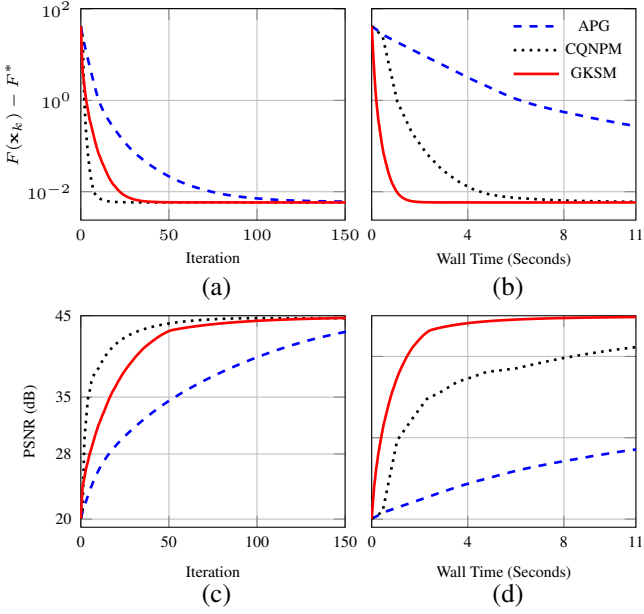
\begin{figure}[t]
	\centering
	\begin{tikzpicture}
  \pgfmathsetmacro{\minybrainOne}{-13.73} 
  \begin{axis}[
      name=CostIter,
     at={(0,0)},
    anchor=south west,
	width=0.28\textwidth,
 	xmin=0, xmax=150,
      xlabel={Iteration},
      ylabel={$F(\uvx_k)-F^*$},
      tick label style={font=\fontsize{6}{6.5}\selectfont},
      label style={font=\fontsize{6}{6.5}\selectfont},
      ylabel style = {yshift=-5pt},
      xlabel style = {yshift=2pt},
      grid=both,
      ymode=log,
      legend style={
  at={(0.5,1),
  font=\fontsize{6}{6.5}\selectfont},
  anchor=north west,
  draw=none,          
  fill=none,          
  cells={align=left},  
  }
  ]

     \addplot [blue,dashed,line width=1pt] table [x expr=\coordindex,y expr=\thisrowno{1}-\minybrainOne] {./figs/MRI/Spiral/DeepBrain1/GradDenoiser_FISTA_LL20/lst_cost_time.txt};

    \addplot [black,dotted,line width=1pt] table [x expr=\coordindex, y expr=\thisrowno{1}-\minybrainOne] {./figs/MRI/Spiral/DeepBrain1/GradDenoiser_QNP20/lst_cost_time.txt};
    
%

    \addplot [red,line width=1pt] table [x expr=\coordindex, y expr=\thisrowno{1}-\minybrainOne] {./figs/MRI/Spiral/DeepBrain1/GradDenoiser_NonGKS20K200RestartFalse/lst_cost_time.txt};

  \end{axis}
  \node[anchor=north west, xshift= 42pt, yshift=-18pt] at (CostIter.south west) {(a)};

   \begin{axis}[
      name=CostTime,
        at={(CostIter.south east)},
    anchor=south west,
	width=0.28\textwidth,
	xshift=10pt,
 xtick={0, 4, 8, 11},
xticklabels={0, 4, 8, 11},
yticklabels={},
 	xmin=0, xmax=11,
      xlabel={Wall Time (Seconds)},
      tick label style={font=\fontsize{6}{6.5}\selectfont},
      label style={font=\fontsize{6}{6.5}\selectfont},
      ylabel style = {yshift=-5pt},
      xlabel style = {yshift=2pt},
      grid=both,
      ymode=log,
      legend style={
  at={(0.5,1),
  font=\fontsize{6}{6.5}\selectfont},
  anchor=north west,
  draw=none,          
  fill=none,          
  cells={align=left},  
  }
  ]

     \addplot [blue,dashed,line width=1pt] table [x index=0,y expr=\thisrowno{1}-\minybrainOne] {./figs/MRI/Spiral/DeepBrain1/GradDenoiser_FISTA_LL20/lst_cost_time.txt};
       \addlegendentry{APG}

  \addplot [black,dotted,line width=1pt] table [x index=0, y expr=\thisrowno{1}-\minybrainOne] {./figs/MRI/Spiral/DeepBrain1/GradDenoiser_QNP20/lst_cost_time.txt};
     \addlegendentry{CQNPM}

    \addplot [red,line width=1pt] table [x index=0, y expr=\thisrowno{1}-\minybrainOne] {./figs/MRI/Spiral/DeepBrain1/GradDenoiser_NonGKS20K200RestartFalse/lst_cost_time.txt};
      \addlegendentry{\PropM}
  \end{axis}
  \node[anchor=north west, xshift= 42pt, yshift=-18pt] at (CostTime.south west) {(b)};

 \begin{axis}[
      name=PSNRIter,
     at={(CostIter.south)},
    anchor=north,
	width=0.28\textwidth,
	yshift = -36pt,
 xtick={0, 50, 100, 150},
  xticklabels={0, 50, 100, 150},
   ytick={20, 28, 35, 45},
  yticklabels={20, 28, 35, 45},
  	xmin=0, xmax=150,
    ymin=19, ymax=45,
      xlabel={Iteration},
      ylabel={PSNR (dB)},
      tick label style={font=\fontsize{6}{6.5}\selectfont},
      label style={font=\fontsize{6}{6.5}\selectfont},
      grid=both,
      legend style={
  at={(0.67,0.36),
  font=\fontsize{5}{5.5}\selectfont},
  anchor=north west,
  draw=none,          
  fill=none,          
  cells={align=left},  
  }
  ]

     \addplot [blue,dashed,line width=1pt] table [y index=1, x expr=\coordindex] {./figs/MRI/Spiral/DeepBrain1/GradDenoiser_FISTA_LL20/lst_psnr_time.txt};

    \addplot [black,dotted,line width=1pt] table [y index=1, x expr=\coordindex] {./figs/MRI/Spiral/DeepBrain1/GradDenoiser_QNP20/lst_psnr_time.txt};
     
    \addplot [red,line width=1pt] table [y index=1, x expr=\coordindex] {./figs/MRI/Spiral/DeepBrain1/GradDenoiser_NonGKS20K200RestartFalse/lst_psnr_time.txt};

  \end{axis}
  \node[anchor=north west, xshift= 42pt, yshift=-18pt] at (PSNRIter.south west) {(c)};

  \begin{axis}[
      name=PSNRTime,
     at={(CostTime.south)},
    anchor=north,
    yshift = -36pt,
	width=0.28\textwidth,
 xtick={0, 4, 8, 11},
xticklabels={0, 4, 8, 11},
  yticklabels={},
  	xmin=0, xmax=11,
    ymin=19, ymax=45,
     xlabel={Wall Time (Seconds)},
      tick label style={font=\fontsize{6}{6.5}\selectfont},
      label style={font=\fontsize{6}{6.5}\selectfont},
      grid=both,
      legend style={
  at={(0.67,0.36),
  font=\fontsize{5}{5.5}\selectfont},
  anchor=north west,
  draw=none,          
  fill=none,          
  cells={align=left},  
  }
  ]

       \addplot [blue,dashed,line width=1pt] table [y index=1, x index=0] {./figs/MRI/Spiral/DeepBrain1/GradDenoiser_FISTA_LL20/lst_psnr_time.txt};

    \addplot [black,dotted,line width=1pt] table [y index=1, x index=0] {./figs/MRI/Spiral/DeepBrain1/GradDenoiser_QNP20/lst_psnr_time.txt};
     
    \addplot [red,line width=1pt] table [y index=1, x index=0] {./figs/MRI/Spiral/DeepBrain1/GradDenoiser_NonGKS20K200RestartFalse/lst_psnr_time.txt};
    
  \end{axis}
  \node[anchor=north west, xshift= 42pt, yshift=-18pt] at (PSNRTime.south west) {(d)};
\end{tikzpicture}
\caption{Comparison of different methods with spiral acquisition on the brain $1$ image
for $\varepsilon = 5\times10^{-3}$.
(a), (b): cost values versus iteration and wall time;
(c), (d): PSNR values versus iteration and wall time.}
\label{fig:SpiralBrain1CostPSNR}
\end{figure}

%% file: figs/SpiralBrain1RecoIm.tex
\begin{figure*}[t]
	\centering

   \begin{tikzpicture}

   \begin{axis}[at={((0,0))},anchor = north west,
    xmin = 0,xmax = 250,ymin = 0,ymax = 70, width=1\textwidth,
        scale only axis,
        enlargelimits=false,
        axis line style={draw=none},
        tick style={draw=none},
        axis equal image,
        xticklabels={,,},yticklabels={,,},
       ]

    \node[inner sep=0pt, anchor = south west] (FISTA_50) at (0,0) {\includegraphics[width=0.16\textwidth]{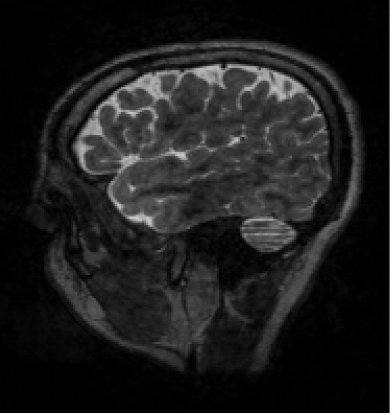}};
   \node at (10,40) {\color{white} $\text{iter.}=50$};
   \node at (35,40) {\color{white} APG};
    \node at (8.5,3) {\color{red} $34.5$dB};
     \node at (33,3) {\color{red} $0.8158$};

    \node[inner sep=0pt, anchor = west] (QNP50) at (FISTA_50.east) {\includegraphics[ width=0.16\textwidth]{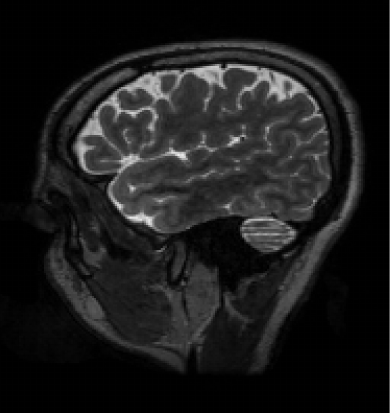}};
      \node at (71,40) {\color{white} CQNPM};
      \node at (49,3) {\color{red} $44.0$dB};
	\node at (72.5,3) {\color{red} $0.9760$};
 
     \node[inner sep=0pt, anchor = west] (NKSM50) at (QNP50.east) {\includegraphics[ width=0.16\textwidth]{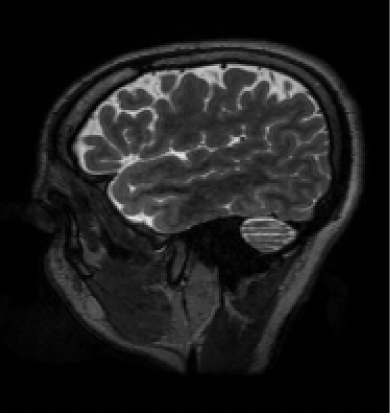}};
      \node at (112,40) {\color{white} \PropM};
      \node at (89,3) {\color{red} $43.0$dB};
	  \node at (113,3) {\color{red} $0.9724$};

    \node[inner sep=3pt, anchor = west] (FISTA_100) at (NKSM50.east) {\includegraphics[ width=0.16\textwidth]{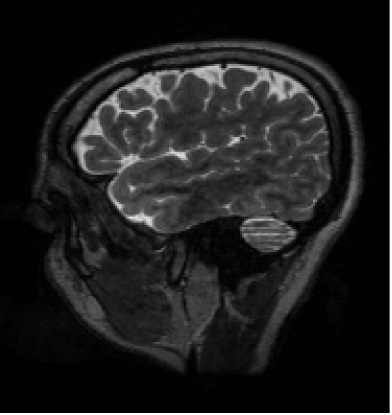}};
  \node at (126,40) {\color{white} $100$};
  \node at (157,40) {\color{white} APG};
  \node at (130,3) {\color{red} $39.9$dB};
   \node at (155,3) {\color{red} $0.9417$};

    \node[inner sep=-3pt, anchor = west] (QNP100) at (FISTA_100.east) {\includegraphics[width=0.16\textwidth]
     {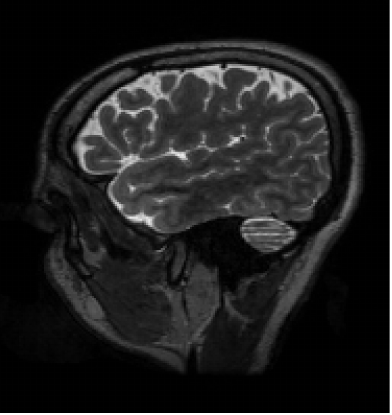}};
     
     \node at (193,39) {\color{white} CQNPM};
      \node at (170,3) {\color{red} $44.7$dB};
      \node at (194,3) {\color{red} $0.9775$}; 
    
   \node[inner sep=3pt, anchor = west] (NKSM100) at (QNP100.east) {\includegraphics[ width=0.16\textwidth]{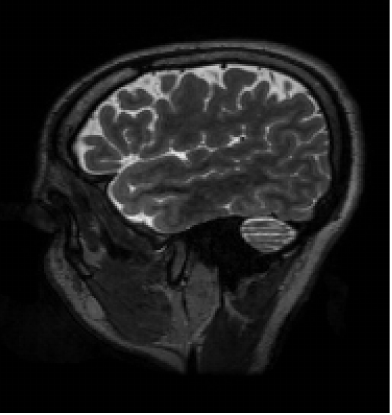}};
     
     \node at (234,40) {\color{white} \PropM};
   \node at (210,3) {\color{red} $44.4$dB};
 \node at (234.7,3) {\color{red} $0.9770$};
\end{axis}

 \begin{axis}[at={(FISTA_50.south west)},anchor = north west,
     xmin = 0,xmax = 250,ymin = 0,ymax = 70, width=1\textwidth,
         scale only axis,
         enlargelimits=false,
         yshift=2cm,
        axis line style={draw=none},
        tick style={draw=none},
         axis equal image,
         xticklabels={,,},yticklabels={,,},
         ylabel style={yshift=-0.2cm,xshift=-0.6cm},
        ]

     \node[inner sep=0pt, anchor = south west] (FISTA_50) at (0,0) {\includegraphics[width=0.16\textwidth]{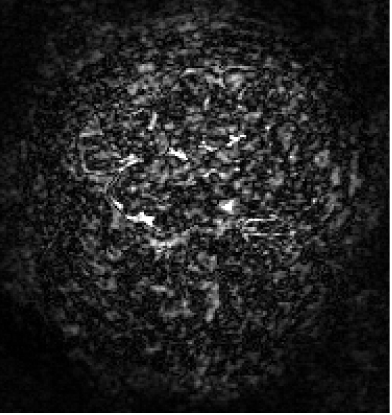}};
    \node at (6,3) {\Large \color{red} $8 \times$};
    \node[inner sep=0pt, anchor = west] (QNP_err_50) at (FISTA_50.east) {\includegraphics[ width=0.16\textwidth]{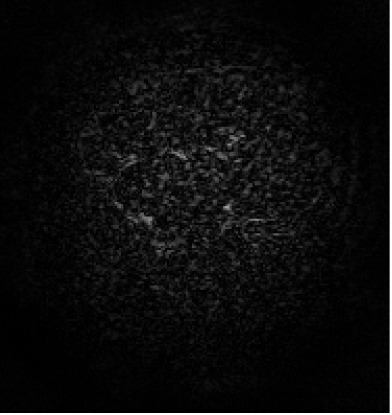}};
      \node[inner sep=0pt, anchor = west] (NKSM_err_50) at (QNP_err_50.east) {\includegraphics[ width=0.16\textwidth]{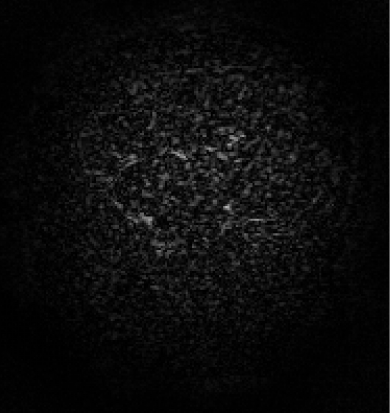}};

    \node[inner sep=3pt, anchor = west] (FISTA_err_100) at (NKSM_err_50.east) {\includegraphics[ width=0.16\textwidth]{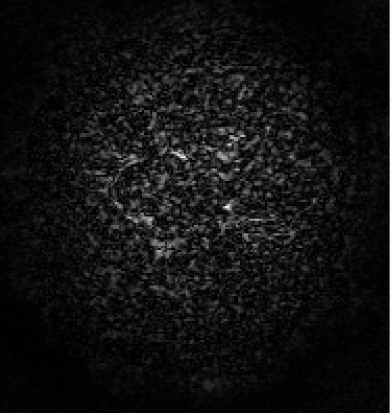}};  
     \node[inner sep=-3pt, anchor = west] (QNP_err_100) at (FISTA_err_100.east) {\includegraphics[ width=0.16\textwidth]
     {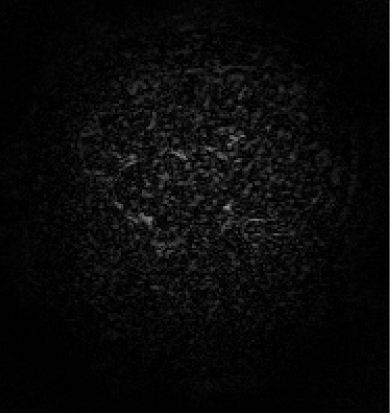}};
   \node[inner sep=3pt, anchor = west] (NKSM_err_100) at (QNP_err_100.east) {\includegraphics[ width=0.16\textwidth]{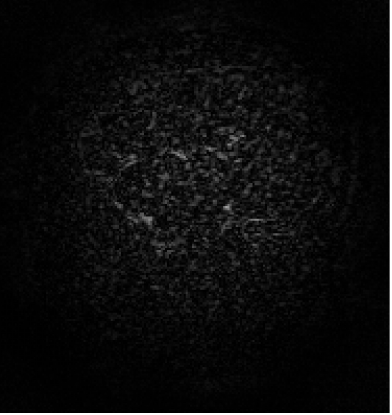}};
   
 \end{axis}
 
\end{tikzpicture} 

\caption{First row: the reconstructed brain $1$ images of each method at $50$th and $100$th iterations with spiral acquisition. The PSNR (\MRcb{respectively, SSIM}) values are labeled at the left (\MRcb{respectively, right}) bottom corner of each image. Second row: the associated error maps ($8 \times$) of the reconstructed images.}
\label{fig:SpiralBrain1:visual}
\end{figure*}

%% file: figs/RadialKnee7CostPSNR.tex
\begin{figure}[t]
	\centering
	\begin{tikzpicture}
  \pgfmathsetmacro{\minybrainOne}{0.01} 
  \begin{axis}[
      name=CostIter,
     at={(0,0)},
    anchor=south west,
	width=0.28\textwidth,
 xtick={0, 25,50,75, 100},
  xticklabels={0, 25,50,75, 100},
 	xmin=0, xmax=100,
      xlabel={Iteration},
      ylabel={$F(\uvx_k)-F^*$},
      tick label style={font=\fontsize{6}{6.5}\selectfont},
      label style={font=\fontsize{6}{6.5}\selectfont},
      ylabel style = {yshift=-5pt},
      xlabel style = {yshift=2pt},
      grid=both,
      ymode=log,
      legend style={
  at={(0.52,1),
  font=\fontsize{6}{6.5}\selectfont},
  anchor=north west,
  draw=none,          
  fill=none,          
  cells={align=left},  
  }
  ]
    
    \addplot [blue,dashed,line width=1pt] table [x expr=\coordindex,y expr=\thisrowno{1}-\minybrainOne] {./figs/MRI/Radial/Knee7/GradDenoiser_FISTA_LL20/lst_cost_time.txt};

    \addplot [black,dotted,line width=1pt] table [x expr=\coordindex, y expr=\thisrowno{1}-\minybrainOne] {./figs/MRI/Radial/Knee7/GradDenoiser_QNP20/lst_cost_time.txt};
     
    \addplot [red,line width=1pt] table [x expr=\coordindex, y expr=\thisrowno{1}-\minybrainOne] {./figs/MRI/Radial/Knee7/GradDenoiser_NonGKS20K200RestartFalse/lst_cost_time.txt};
  
  \end{axis}
  
  \node[anchor=north west, xshift= 42pt, yshift=-18pt] at (CostIter.south west) {(a)};

   \begin{axis}[
      name=CostTime,
        at={(CostIter.south east)},
    anchor=south west,
	width=0.28\textwidth,
	xshift=10pt,
 xtick={0, 1.5,3,4.5, 6},
  xticklabels={0, 1.5,3,4.5, 6},
ytick={0.1, 1, 100},
yticklabels={},
 	xmin=0, xmax=6,
      xlabel={Wall Time (Seconds)},
      tick label style={font=\fontsize{6}{6.5}\selectfont},
      label style={font=\fontsize{6}{6.5}\selectfont},
      ylabel style = {yshift=-5pt},
      xlabel style = {yshift=2pt},
      grid=both,
      ymode=log,
      legend style={
  at={(0.4,0.6),
  font=\fontsize{6}{6.5}\selectfont},
  anchor=north west,
  draw=none,          
  fill=none,          
  cells={align=left},  
  }
  ]

   \addplot [blue,dashed,line width=1pt] table [x index=0,y expr=\thisrowno{1}-\minybrainOne] {./figs/MRI/Radial/Knee7/GradDenoiser_FISTA_LL20/lst_cost_time.txt};
     \addlegendentry{APG}

  \addplot [black,dotted,line width=1pt] table [x index=0, y expr=\thisrowno{1}-\minybrainOne] {./figs/MRI/Radial/Knee7/GradDenoiser_QNP20/lst_cost_time.txt};
     \addlegendentry{CQNPM}

    \addplot [red,line width=1pt] table [x index=0, y expr=\thisrowno{1}-\minybrainOne] {./figs/MRI/Radial/Knee7/GradDenoiser_NonGKS20K200RestartFalse/lst_cost_time.txt};

\addlegendentry{\PropM}
  \end{axis}
  \node[anchor=north west, xshift= 42pt, yshift=-18pt] at (CostTime.south west) {(b)};
  
\begin{axis}[
      name=PSNRIter,
     at={(CostIter.south)},
    anchor=north,
	width=0.28\textwidth,
	yshift = -36pt,
 xtick={0,25, 50,75, 100},
  xticklabels={0, 25,50,75, 100},
  ytick={18, 28, 35, 43},
 yticklabels={18, 28, 35, 43},
  	xmin=0, xmax=100,
  	ymax=43.5,
    ymin=18, 
      xlabel={Iteration},
      ylabel={PSNR (dB)},
      tick label style={font=\fontsize{6}{6.5}\selectfont},
      label style={font=\fontsize{6}{6.5}\selectfont},
      grid=both,
      legend style={
  at={(0.67,0.36),
  font=\fontsize{5}{5.5}\selectfont},
  anchor=north west,
  draw=none,          
  fill=none,          
  cells={align=left},  
  }
  ]
    
    \addplot [blue,dashed,line width=1pt] table [y index=1, x expr=\coordindex] {./figs/MRI/Radial/Knee7/GradDenoiser_FISTA_LL20/lst_psnr_time.txt};

    \addplot [black,dotted,line width=1pt] table [y index=1, x expr=\coordindex] {./figs/MRI/Radial/Knee7/GradDenoiser_QNP20/lst_psnr_time.txt};
     
    \addplot [red,line width=1pt] table [y index=1, x expr=\coordindex] {./figs/MRI/Radial/Knee7/GradDenoiser_NonGKS20K200RestartFalse/lst_psnr_time.txt};

  \end{axis}
  \node[anchor=north west, xshift= 42pt, yshift=-18pt] at (PSNRIter.south west) {(c)};

  \begin{axis}[
      name=PSNRTime,
     at={(CostTime.south)},
    anchor=north,
    yshift = -36pt,
	width=0.28\textwidth,
  yticklabels={},
   xtick={0, 1.5,3,4.5, 6},
  xticklabels={0, 1.5,3,4.5, 6},
  	xmin=0, xmax=6,
    ymin=18, ymax=43,
     xlabel={Wall Time (Seconds)},
      tick label style={font=\fontsize{6}{6.5}\selectfont},
      label style={font=\fontsize{6}{6.5}\selectfont},
      grid=both,
      legend style={
  at={(0.67,0.36),
  font=\fontsize{5}{5.5}\selectfont},
  anchor=north west,
  draw=none,          
  fill=none,          
  cells={align=left},  
  }
  ]
    
 \addplot [blue,dashed,line width=1pt] table [y index=1, x index=0] {./figs/MRI/Radial/Knee7/GradDenoiser_FISTA_LL20/lst_psnr_time.txt};

    \addplot [black,dotted,line width=1pt] table [y index=1, x index=0] {./figs/MRI/Radial/Knee7/GradDenoiser_QNP20/lst_psnr_time.txt};
     
    \addplot [red,line width=1pt] table [y index=1, x index=0] {./figs/MRI/Radial/Knee7/GradDenoiser_NonGKS20K200RestartFalse/lst_psnr_time.txt};

  \end{axis}
  \node[anchor=north west, xshift= 42pt, yshift=-18pt] at (PSNRTime.south west) {(d)};
\end{tikzpicture}
\caption{Comparison of different methods with radial acquisition on the knee $1$ image
for $\varepsilon = 6\times10^{-3}$. (a), (b): cost values versus iteration and wall time; (c), (d): PSNR values versus iteration and wall time.
}
\label{fig:RadialKnee1CostPSNR}
\end{figure}
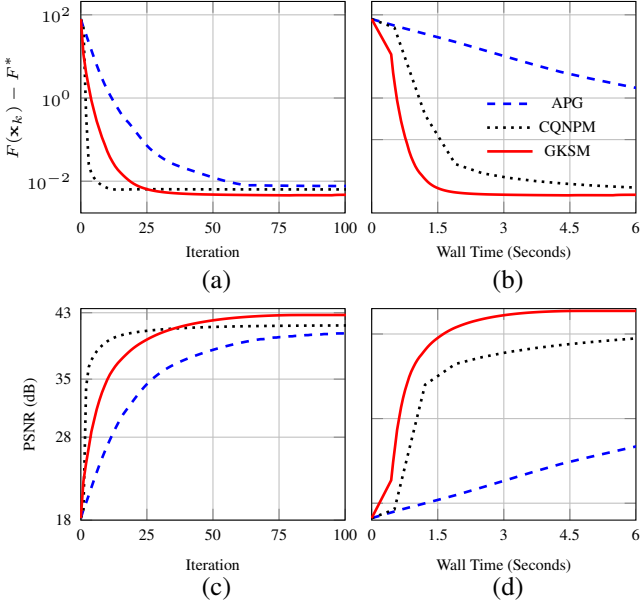

%% file: figs/RadialKnee7RecoIm.tex
\begin{figure*}[t]
	\centering

   \begin{tikzpicture}

   \begin{axis}[at={((0,0))},anchor = north west,
    xmin = 0,xmax = 250,ymin = 0,ymax = 70, width=1\textwidth,
        scale only axis,
        enlargelimits=false,
        axis line style={draw=none},
        tick style={draw=none},
        axis equal image,
        xticklabels={,,},yticklabels={,,},
       ]

   \node[inner sep=0pt, anchor = south west] (FISTA_50) at (0,0) {\includegraphics[width=0.16\textwidth]{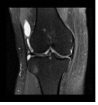}};
   \node at (10,40) {\color{white} $\text{iter.}=50$};
   \node at (34.5,40) {\color{white} APG};
    \node at (8.2,3.5) {\color{red} $38.5$dB};
     \node at (32,3.5) {\color{red} $0.9028$};

    \node[inner sep=-1pt, anchor = west] (QNP50) at (FISTA_50.east) {\includegraphics[ width=0.16\textwidth]{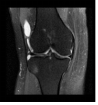}};
      \node at (70,40) {\color{white} CQNPM};
      \node at (48,3.5) {\color{red} $41.3$dB};
	  \node at (71,3.5) {\color{red} $0.9422$};
 
     \node[inner sep=0pt, anchor = west] (NKSM50) at (QNP50.east) {\includegraphics[ width=0.16\textwidth]{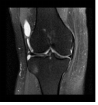}};
      \node at (111.5,40) {\color{white} \PropM};
      \node at (88,3.5) {\color{red} $42.0$dB};
	  \node at (111,3.5) {\color{red} $0.9389$};
      
    \node[inner sep=2pt, anchor = west] (FISTA_100) at (NKSM50.east) {\includegraphics[ width=0.16\textwidth]{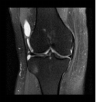}};
  \node at (124,40) {\color{white} $100$};
  \node at (154,40) {\color{white} APG};
  \node at (128,3.5) {\color{red} $40.5$dB};
   \node at (152.5,3.5) {\color{red} $0.9382$};
    
     \node[inner sep=-4pt, anchor = west] (QNP100) at (FISTA_100.east) {\includegraphics[width=0.16\textwidth]
     {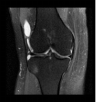}};
     
     \node at (190,40) {\color{white} CQNPM};
      \node at (168,3.5) {\color{red} $41.5$dB};
     \node at (191,3.5) {\color{red} $0.9377$};
    
   \node[inner sep=3pt, anchor = west] (NKSM100) at (QNP100.east) {\includegraphics[ width=0.16\textwidth]{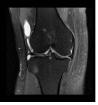}};
     
     \node at (231,40) {\color{white} \PropM};
   	\node at (208,3.5) {\color{red} $42.7$dB};
	\node at (231,3.5) {\color{red} $0.9102$};

\end{axis}

 \begin{axis}[at={(FISTA_50.south west)},anchor = north west,
     xmin = 0,xmax = 250,ymin = 0,ymax = 70, width=1\textwidth,
         scale only axis,
         enlargelimits=false,
         yshift=2.05cm,
        axis line style={draw=none},
        tick style={draw=none},
         axis equal image,
         xticklabels={,,},yticklabels={,,},
         ylabel style={yshift=-0.2cm,xshift=-0.6cm},
        ]

\node[inner sep=0pt, anchor = south west] (FISTA_50) at (0,0) {\includegraphics[width=0.16\textwidth]{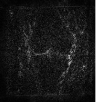}};
    \node at (5,4) {\Large \color{red} $8 \times$};
    
    \node[inner sep=-1pt, anchor = west] (QNP_err_50) at (FISTA_50.east) {\includegraphics[ width=0.16\textwidth]{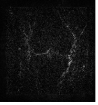}};
      \node[inner sep=0pt, anchor = west] (NKSM_err_50) at (QNP_err_50.east) {\includegraphics[ width=0.16\textwidth]{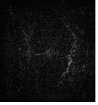}};

    \node[inner sep=2pt, anchor = west] (FISTA_err_100) at (NKSM_err_50.east) {\includegraphics[ width=0.16\textwidth]{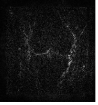}};  
    
     \node[inner sep=-4pt, anchor = west] (QNP_err_100) at (FISTA_err_100.east) {\includegraphics[ width=0.16\textwidth]
     {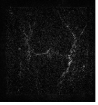}};
   \node[inner sep=3pt, anchor = west] (NKSM_err_100) at (QNP_err_100.east) {\includegraphics[ width=0.16\textwidth]{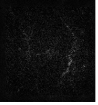}};      
 \end{axis}
 
\end{tikzpicture} 

\caption{First row: the reconstructed knee $1$ images of each method at $50$th and $100$th iterations with radial acquisition. The PSNR (\MRcb{respectively, SSIM}) values are labeled at the left (\MRcb{respectively, right}) bottom corner of each image. Second row: the associated error maps ($8\times$) of the reconstructed images.}
\label{fig:RadialKnee1:visual}
\end{figure*}

%% file: figs/SpiralBrainConvVal.tex
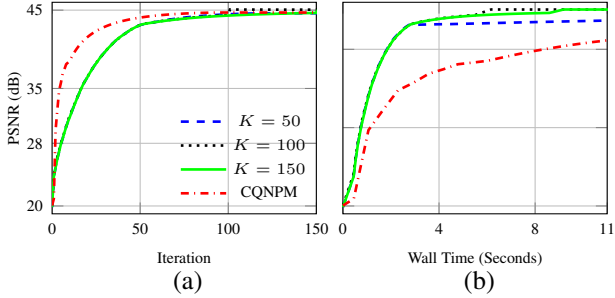
\begin{figure}[t]
	\centering
	\begin{tikzpicture}
  \pgfmathsetmacro{\minybrainOne}{-13.73} 
  
   \begin{axis}[
      name=PSNRIter,
     at={(0,0)},
    anchor=north,
	width=0.28\textwidth,
 xtick={0, 50, 100, 150},
  xticklabels={0, 50, 100, 150},
   ytick={20, 28, 35, 45},
  yticklabels={20, 28, 35, 45},
  	xmin=0, xmax=150,
    ymin=19, ymax=46,
      xlabel={Iteration},
      ylabel={PSNR (dB)},
      tick label style={font=\fontsize{6}{6.5}\selectfont},
      label style={font=\fontsize{6}{6.5}\selectfont},
      ylabel style = {yshift=-5pt},
      grid=both,
      legend style={
  at={(0.45,0.52),
  font=\fontsize{6}{6.5}\selectfont},
  anchor=north west,
  draw=none,          
  fill=none,          
  cells={align=left},  
  }
  ]

\addplot [blue,dashed,line width=1pt] table [y index=1, x expr=\coordindex] {./figs/ablation/lst_psnr_timeK50.txt};
\addlegendentry{$K=50$}

\addplot [black,dotted,line width=1pt] table [y index=1, x expr=\coordindex] {./figs/ablation/lst_psnr_timeK100.txt};
\addlegendentry{$K=100$}

\addplot [green,solid,line width=1pt] table [y index=1, x expr=\coordindex] {./figs/ablation/lst_psnr_timeK150.txt};
\addlegendentry{$K=150$}

\addplot [red,dash dot,line width=1pt] table [y index=1, x expr=\coordindex] {./figs/MRI/Spiral/DeepBrain1/GradDenoiser_QNP20/lst_psnr_time.txt};
\addlegendentry{CQNPM}

  \end{axis}
  \node[anchor=north west, xshift= 42pt, yshift=-18pt] at (PSNRIter.south west) {(a)};

  \begin{axis}[
      name=PSNRTime,
     at={(PSNRIter.south east)},
    anchor=south west,
   xshift=10pt,
	width=0.28\textwidth,
 xtick={0, 4, 8, 11},
xticklabels={0, 4, 8, 11},
  yticklabels={},
  	xmin=0, xmax=11,
    ymin=19, ymax=46,
     xlabel={Wall Time (Seconds)},
      tick label style={font=\fontsize{6}{6.5}\selectfont},
      label style={font=\fontsize{6}{6.5}\selectfont},
      grid=both,
      legend style={
  at={(0.67,0.36),
  font=\fontsize{5}{5.5}\selectfont},
  anchor=north west,
  draw=none,          
  fill=none,          
  cells={align=left},  
  }
  ]

\addplot [blue,dashed,line width=1pt] table [y index=1, x index=0] {./figs/ablation/lst_psnr_timeK50.txt};
 
\addplot [black,dotted,line width=1pt] table [y index=1, x index=0] {./figs/ablation/lst_psnr_timeK100.txt};

\addplot [green,solid,line width=1pt] table [y index=1, x index=0] {./figs/ablation/lst_psnr_timeK150.txt};

\addplot [red,dash dot,line width=1pt] table [y index=1, x index=0] {./figs/MRI/Spiral/DeepBrain1/GradDenoiser_QNP20/lst_psnr_time.txt};

  \end{axis}
  \node[anchor=north west, xshift= 42pt, yshift=-18pt] at (PSNRTime.south west) {(b)};

\end{tikzpicture}
\caption{Comparison of varying $K$ with spiral acquisition on the brain $1$ image. (a), (b): PSNR values versus iteration and wall time.}
\label{fig:SpiralBrain1VarK}
\end{figure}

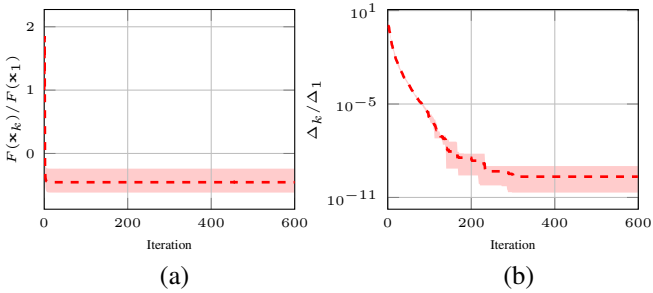
\begin{figure}
    \centering
\begin{tikzpicture}
\pgfplotsset{set layers}
  \begin{axis}[
      name=CostIter,
     at={(0,0)},
     xshift=-20pt,
    anchor=south west,
	width=0.27\textwidth,
 	xmin=0, xmax=600,
      xlabel={Iteration},
      ylabel={$F(\uvx_k)/F(\uvx_1)$},
      tick label style={font=\fontsize{5}{5.5}\selectfont},
      label style={font=\fontsize{5}{5.5}\selectfont},
      ylabel style = {yshift=-5pt},
      xlabel style = {yshift=2pt},
      grid=both,
      legend style={
  at={(0.6,1),
  font=\fontsize{5}{5.5}\selectfont},
  anchor=north west,
  draw=none,          
  fill=none,          
  cells={align=left},  
  }
  ]
  
\addplot[name path=lower,dashed, red,draw=none] table [x expr=\coordindex, y expr=\thisrowno{0}] {./figs/ablation/Conv/K600SpiralBrain_cost_min.txt};
  
\addplot[name path=upper,dashed,blue,draw=none] table [x expr=\coordindex, y expr=\thisrowno{0}] {./figs/ablation/Conv/K600SpiralBrain_cost_max.txt};  

\tikzfillbetween[of=lower and upper,on layer=axis background]{fill=red!20};

\addplot [red,dashed,line width=1pt] table [x expr=\coordindex, y expr=\thisrowno{0}] {./figs/ablation/Conv/K600SpiralBrain_cost_mean.txt};
\end{axis}
\node[anchor=north west, xshift= 20pt, yshift=-18pt] at (CostIter.south west) {(a)};
  
  \begin{axis}[
      name=IterNormIter,
        at={(CostIter.south east)},
    anchor=south west,
    ylabel = {$\Delta_k/\Delta_1$},
	width=0.27\textwidth,
	xshift=15pt,
 	xmin=0, xmax=600,
      xlabel={Iteration},
      tick label style={font=\fontsize{5}{5.5}\selectfont},
      label style={font=\fontsize{5}{5.5}\selectfont},
      ylabel style = {yshift=-5pt},
      xlabel style = {yshift=2pt},
      grid=both,
      ymode=log,
      legend style={
  at={(0.67,1),
  font=\fontsize{5}{5.5}\selectfont},
  anchor=north west,
  draw=none,          
  fill=none,          
  cells={align=left},  
  }
  ]
    
 \addplot[name path=lower,dashed, red,draw=none] table [x expr=\coordindex, y expr=\thisrowno{0}] {./figs/ablation/Conv/K600SpiralBrain_iter_norm_min.txt};
  
\addplot[name path=upper,dashed,blue,draw=none] table [x expr=\coordindex, y expr=\thisrowno{0}] {./figs/ablation/Conv/K600SpiralBrain_iter_norm_max.txt};  

\tikzfillbetween[of=lower and upper,on layer=axis background]{fill=red!20};

\addplot [red,dashed,line width=1pt] table [x expr=\coordindex, y expr=\thisrowno{0}] {./figs/ablation/Conv/K600SpiralBrain_iter_norm_mean.txt};
  \end{axis}
\node[anchor=north west, xshift= 55pt, yshift=-18pt] at (IterNormIter.south west) {(b)};
  
  \end{tikzpicture}
    \caption{(a) Averaged cost values (a) and $\Delta_k/\Delta_1$ (b) versus iteration for \PropM. The shaded region of each curve represents the range of the cost values and $\Delta_k$
    across six brain test images with spiral acquisition.}
    \label{fig:ConvVal}
\end{figure}